\documentclass{aa}                   
\usepackage{graphicx}                
\usepackage{txfonts}                 

\newcommand{\Halpha}{H$\alpha$}
\newcommand{\kms}{km\,s$^{-1}$}
\newcommand{\ms}{m\,s$^{-1}$}

\begin{document}

\title{Binary-induced magnetic activity?\thanks{Based on data obtained with the STELLA robotic
telescope in Tenerife, an AIP facility jointly operated by AIP and
IAC, and the Potsdam Automatic Photoelectric Telescopes (APT) in
Arizona, jointly operated by AIP and Fairborn Observatory.}}
\subtitle{Time-series echelle spectroscopy and photometry of
HD\,123351 = CZ~CVn}

\author{K. G.~Strassmeier\inst{1}, T. A.~Carroll\inst{1}, M.~Weber\inst{1}, T.~Granzer\inst{1},
J.~Bartus\inst{1,2}, K.~Ol\'ah\inst{2} \and J.~B.~Rice\inst{3}}

\offprints{K. G. Strassmeier}

\institute{Astrophysical Institute Potsdam, An der Sternwarte 16,
D-14482 Potsdam, Germany, \email{KStrassmeier@aip.de}, \and
Konkoly Observatory, H-1525 Budapest, P.O.Box 67, Hungary,
\email{olah@konkoly.hu} \and Department of Physics, Brandon
University, Brandon, Manitoba R7A 6A9, Canada
\email{rice@BrandonU.ca} }

\date{Received ... ; accepted ...}

\abstract{Multi-wavelength time-series observations with high
cadence and long duration are needed to resolve and understand the
many variations of magnetically active late-type stars, which is
an approach often used to observe the Sun.}{We present a first and
detailed study of the bright and active K0IV-III star
HD~123351.}{We acquired a total of 955 high-resolution STELLA
echelle spectra during the years 2006--2010 and a total of 2,260
photometric $VI_C$ data points during 1998--2010. These data are
complemented by some spectra from CFHT and KPNO.}{The star is
found to be a single-lined spectroscopic binary with a period of
147.8919$\pm$0.0003 days and a large eccentricity of
$e$=0.8086$\pm$0.0001. The rms of the orbital solution is just
47~\ms , making it the most precise orbit ever obtained for an
active binary system. The rotation period is constrained from
long-term photometry to be 58.32$\pm$0.01~days. It shows that
HD~123351 is a very asynchronous rotator, rotating five times
slower than the expected pseudo-synchronous value. Two spotted
regions persisted throughout the 12 years of our observations. We
interpret them as active longitudes on a differentially rotating
surface with a $\Delta P/P$ of 0.076. Four years of \Halpha ,
Ca\,{\sc ii} H\&K and He\,{\sc i}~D3 monitoring identifies the
same main periodicity as the photometry but dynamic spectra also
indicate that there is an intermittent dependence on the orbital
period, in particular for Ca\,{\sc ii} H\&K in 2008. Line-profile
inversions of a pair of Zeeman sensitive/insensitive iron lines
yield an average surface magnetic-flux density of 542$\pm$72~G.
The time series for 2008 is modulated by the stellar rotation as
well as the orbital motion, such that the magnetic flux is
generally weaker during times of periastron and that the
chromospheric emissions vary in anti-phase with the magnetic flux.
We also identify a broad and asymmetric lithium line profile and
measure an abundance of $\log n({\rm Li})=1.70\pm 0.05$. The
star's position in the H-R diagram indicates a mass of
1.2$\pm$0.1~M$_\odot$ and an age of 6-7~Gyr.}{We interpret the
anti-phase relation of the magnetic flux with the chromospheric
emissions as evidence that there are two magnetic fields present
at the same time, a localized surface magnetic field associated
with spots and a global field that is oriented towards the
(low-mass) secondary component. We suggest that the inter-binary
field is responsible for the magnetic-flux dilution at periastron.
It is also likely to be responsible for the unexpected slow and
asynchronous rotation of the primary star.}


\authorrunning{K. G. Strassmeier et al.}

\maketitle

\section{Introduction}

Cool main-sequence stars in binaries appear to be magnetically
very active when the orbital period is as short as a few days.
Evolved components in binaries are suspected to display
photospheric spots and strong chromospheric emission at
significantly longer periods, say, 20 days. This has been
generally interpreted as meaning that there is a stellar dynamo
that needs a convective envelope with a long-enough turn-over time
and short-enough rotation to function efficiently (Durney \&
Latour \cite{dur:lat}). The existence of active longitudes in
tidally interacting binaries is part of the proof of this concept
(see, e.g., Berdyugina \cite{berd2007}). The observational
evidence is less clear for stars in a binary where the rotation
and the orbital revolution is not synchronized or, even more
complex, if having additionally a highly eccentric orbit (Hall
\cite{h91}, Strassmeier \cite{eadn}). A binary system with an
active component and all of above properties, in particular with a
high eccentricity, may allow us to prove or disprove the
speculation that a dynamo would operate more efficiently at
periastron when the stars are closest, as would be the case for a
dynamo influenced by tidal forces (e.g. Schrijver \& Zwaan
\cite{sch:zwa}). Such a binary system is presented and analyzed in
the present paper.

The star HD\,123351 ($\alpha$ = 14$^h$06$^m$26$^s$, $\delta$ =
30$\degr$50\arcmin47\arcsec\ [2000], $V\approx7\fm6$) was found to
exhibit strong Ca\,{\sc ii} H\&K emission as part of the
Vienna-KPNO Doppler imaging candidate survey (Strassmeier et al.
\cite{ss}). The survey spectra allowed us to determine Ca\,{\sc
ii} H\&K and \Halpha\ emission-line fluxes, a projected rotational
velocity of 5.2$\pm$2~\kms , a logarithmic lithium abundance of
1.9, and provided two radial velocities 160 days apart that
basically agreed within their errors and thus were implied that we
had observed a single star. To our knowledge there had been no
evidence in the literature that the star is a spectroscopic or a
visual binary despite of its relative brightness. Our new data in
the present paper indicate the star is a single-lined
spectroscopic binary with a very eccentric orbit.

The light variability appears to have been first detected by the
\emph{Hipparcos} satellite which led Kazarovets et al.
(\cite{ibvs}) to include HD\,123351 (HIP\,68904) in the 74th name
list of variable stars as CZ~CVn. Simbad identifies it as a
``pulsating star'' with evidence neither for nor against it. The
\emph{Hipparcos} light curve was analyzed a few years later by
Koen \& Eyer (\cite{koe:eye}), who found a photometric period of
59.4 days with an amplitude of 0\fm028. A year earlier we had
published our first three years of Automatic Photoelectric
Telescope (APT) $VI$ data from 1998-2000 (in a paper on APT data
reduction by Granzer et al. \cite{woozi2}) that also clearly
showed the $\approx$60-day periodicity.

The star was detected as an X-ray and EUV source by EUVE and ROSAT
with count rates of 0.88 and 0.021, respectively (Lampton et al.
\cite{lamp}). It is listed in the ROSAT all-sky survey (Schwope et
al.~\cite{rass}) and identified in the Hamburg/RASS bright source
identification catalogue (Zickgraf et al. \cite{zick}). A study of
the X-ray selected Seyfert galaxies by Vaughan et al. (\cite{vau})
listed HD\,123351 in their Galactic sample with a Galactic column
density of 1.25$\times$10$^{20}$~cm$^{-2}$. A study of the local
kinematics of K and M giants with Coravel/Hipparcos/Tycho data
(Famaey et al.~\cite{fam}) included HD\,123351 but did not provide
a new radial-velocity measurement. The new Pulkova radial-velocity
catalog of HIP stars lists HD\,123351 with a single measurement of
$-11.2\pm0.8$~\kms\ (Gontcharov \cite{gont}).

Therefore, the star appeared to be an interesting and rather
bright chromospherically active star and deserved a more detailed
study. In this paper, we present and analyze an extensive
time-series database consisting of high-resolution echelle spectra
obtained as part of the science demonstration of the first of our
two robotic STELLA telescopes in Tenerife, complemented by some
previous KPNO and CFHT spectra, and 12 years of photometric
monitoring with our APT in southern Arizona. In Sect.~\ref{S2}, we
describe our data. In Sect.~3, we focus on the recovery of the
star's basic astrophysical parameters such as the rotational
period, the orbital elements, stellar mass, radius and age,
lithium abundance, Rossby number, and evolutionary status in
general. Section~\ref{S4} presents the time-series analysis of the
photometric and spectroscopic data in order to reconstruct the
stellar surface structure as a function of rotational and orbital
phase. Section~\ref{S5} is an attempt to gather the various
results together. Our summary and conclusions are presented in
Sect.~\ref{S6}.

\begin{figure*}[tb]
\includegraphics[angle=0,width=160mm]{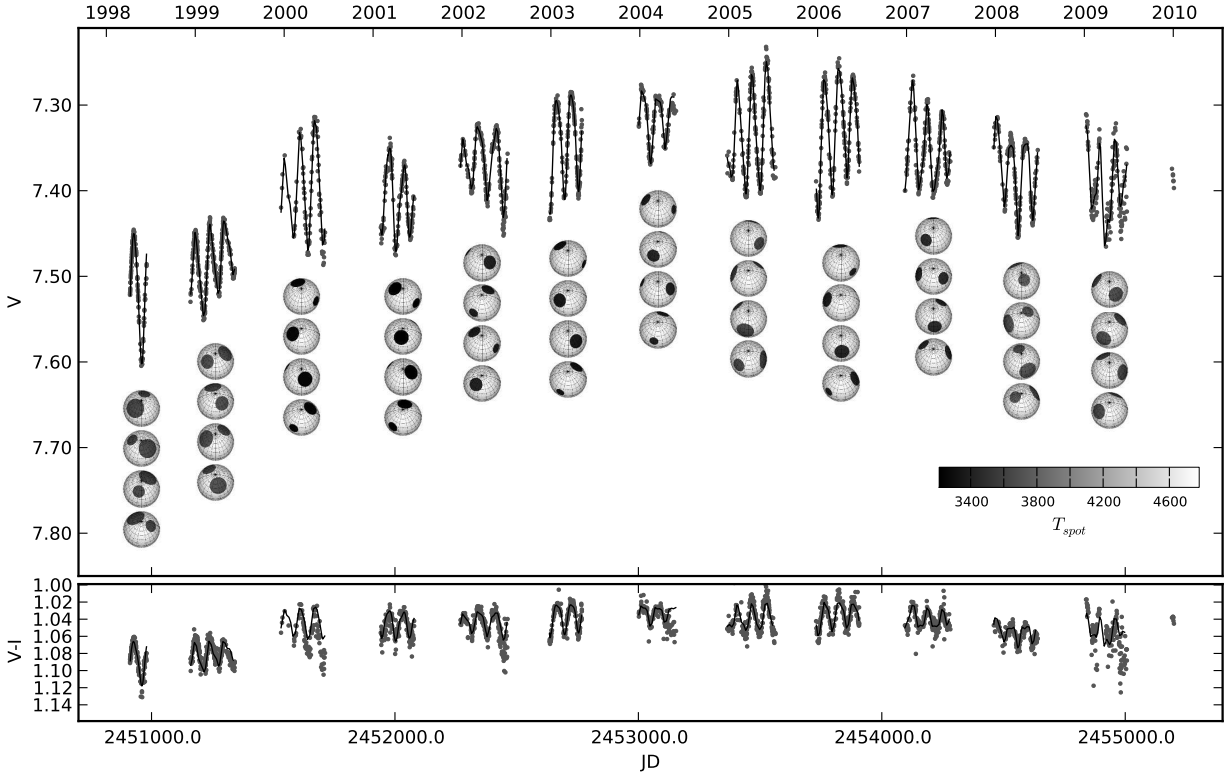}
\caption[ ]{The long-term brightness variations of HD\,123351. The
top panel shows our twelve years of $V$-band photometry (dots),
the lower panel plots the corresponding $V-I$ color index. The
lines are the seasonal fits of the two-spot models from
Sect.~\ref{S4}. These models are plotted here as spherical
projections at four different phases (from top to bottom, 0\degr,
90\degr, 180\degr, 270\degr). The gray scale indicates the spot
temperature. Note the rotational modulation with a period of
$\approx$60~d and a long-term variation in both $V$ and $V-I$ in
the sense that the star appears redder when fainter.
 \label{F1}}
\end{figure*}

\begin{table}[!tbh]
\begin{flushleft}
\caption{Spectroscopic and photometric observing logs. }\label{T1}
\begin{tabular}{lllll}
\hline \noalign{\smallskip}
Observatory & Type & Instr. & Time range & $N$ \\
                        & & & JD 245+     & \\
\noalign{\smallskip}\hline \noalign{\smallskip}
STELLA-I & Echelle spec. & SES & 3912--5213& 955 \\
KPNO CF & Coud\'e spec. & Coud\'e & 0908--1673 & 20 \\
CFHT & Coud\'e spec. & Gecko & 1683--1687 & 4 \\
APT T7 & $VI_C$ & PMT & 0911--5200 & 2260\\
\noalign{\smallskip}\hline
\end{tabular}

\vspace{1mm}Note. $N$ is the total number of individual spectra or
light-curve points. CF denotes the 0.9m Coud\'e Feed telescope at
Kitt Peak. PMT means PhotoMultiplier Tube.
\end{flushleft}
\end{table}

\section{Observations}\label{S2}

\subsection{APT photometry since 1998}

High-precision photometry of HD~123351 was obtained with the
former University of Vienna 0.75m twin automatic photoelectric
telescopes (APTs) at Fairborn Observatory in southern Arizona
(Strassmeier et al. \cite{apt}, Granzer et al.~\cite{woozi2}). The
T7-APT {\sl Amadeus} was used since JD 2,450,900. It achieved an
external precision of 3.2\,mmag in Johnson-Cousins $VI_{\rm C}$. A
total of 2260 $V$ and $I_C$ data points were obtained until the
end of the observing season 2010 and are presented in this paper
in Fig.~\ref{F1}. The integration time was 10~sec in $V$ and
$I_{\rm C}$. All measurements were done differentially with
respect to HD~123929 (G8, $V$=7\fm27, $B-V$=0.85, Oja \cite{oja};
$V_T$=7\fm360, $B-V$=0\fm835, H\o g et al. \cite{hog}). The check
star was HD~123533 (K0). Both stars appear constant to within the
telescope's internal precision. The APT data were transformed to
absolute values with nightly observations of up to 60 standard
stars. For further details of the observing procedure and the APT
data reduction in general, we refer to Granzer et~al.
(\cite{woozi2}).

\subsection{STELLA/SES spectroscopy in 2006--2010}

High-resolution time-series spectroscopy was obtained with the
\emph{STELLA Echelle Spectrograph} (SES) at the robotic 1.2-m
STELLA-I telescope in Tenerife, Spain (Strassmeier et
al.~\cite{stella}, \cite{malaga}). A total of 955 echelle spectra
were acquired over the course of four years. The SES is a
fiber-fed white-pupil echelle spectrograph with a fixed wavelength
format of 388--882\,nm. Despite increasing inter-order gaps in the
red, it records the range 390--720nm continuously. The instrument
is located in a dedicated room on a stabilized optical bench and
is fed by a 12-m long 50$\mu$m Ceram-Optec fibre, corresponding to
an entrance aperture of 1.7\arcsec\ on the sky. This fiber enables
a two-pixel resolution of $R$=55,000. The spectrograph's heart is
a 31 lines per mm R2 grating. Two off-axis parabolic collimators,
one folding mirror, and a prism as a cross disperser transport the
light into the f/2.7 kata-dioptric camera with a 20cm corrector
and a 40cm spherical mirror. The CCD is currently an E2V\,42-40
2048$\times$2048 13.5$\mu$m-pixel device. Its quantum efficiency
is 90\% at 650nm, and 65\% at 400 and 800nm, respectively.
Together with a second-generation CUO controller, the read-out
noise is 3--4 electrons rms at 200~kbyte/s. A closed cycle cooler
keeps the detector cooled to $-130\degr$C.

Integrations on HD~123351 were set to exposure times of between
3600~sec and 1200~sec and achieved a signal-to-noise ratio (S/N)
of between 200:1 to 50:1 per resolution element, respectively,
depending on weather conditions. One spectrum on almost every
clear night was obtained between June 25, 2006 and June 22, 2010.
Numerous radial velocity standards and stellar comparison targets
were also observed with the same set-up. Earlier data in 2006 and
during March-October 2007 had to be shifted by +280~\ms\ because
of a fiber misalignment problem. The time-series velocities are
plotted in Fig.~\ref{F2}. The best rms radial-velocity precision
over the three years of observation was 30~\ms . The time series
also included nightly flat-field exposures, bias readouts, and
Th-Ar comparison-lamp exposures. For details of the echelle data
reduction, we refer to Ritter \& Washuettl (\cite{ritter}) and
Weber et al. (\cite{spie}) and our previous paper on HD~1
(Strassmeier et al.~\cite{hd1}). The data log is summarized in
Table~\ref{T1}.

\subsection{KPNO/CF spectroscopy in 1998 and 2000}

Twenty KPNO observations were obtained with the 0.9-m coud{\'e}
feed (CF) telescope in April 1998 and in March 2000
(Table~\ref{T1}). The F3KB 3k-CCD was used together with
grating~A, the long collimator and a 280~$\mu$m slit. This
configuration allowed a resolving power of $R\approx$28,000 at a
dispersion of 4.81~\kms/pixel. The wavelength coverage was 30~nm
and the spectra were centered at 649~nm. The exposure time was
1200 seconds resulting in an average S/N of 100:1.

All KPNO spectra were reduced and extracted using the Interactive
Reduction and Analysis Facility (IRAF) provided by NOAO. More
details of the standard coud\'e-spectra reduction procedure were
given in previous papers, e.g. by Weber \& Strassmeier
(\cite{ilhya}).

\subsection{CFHT/Gecko spectroscopy in 2000}

Three very high-resolution lithium spectra were obtained
consecutively on one night at CFHT with \emph{Gecko} and an
order-sorting filter in the 9th order at HJD 2,451,687.75. A
fourth integration of 900~sec was taken centered at 644\,nm at
2,451,683.73. Equipped with the CAFE fiber feed module
\emph{Gecko} provided a spectral resolution of 120,000. CAFE feeds
light through a 27m-long 100$\mu$m Ceram-Optec fiber to a
four-slice Bowen-Walraven image slicer before it enters the
collimator. In combination with the 4600$\times$2048
13.5$\mu$m-pixel EEV1 CCD, it gives a 10nm wavelength coverage
centered at 671~nm. An integration time of 600~sec per exposure
gave a peak S/N of 300:1 per resolution element. For a brief
description of the data reduction, we refer to the paper by
Strassmeier \& Rice (\cite{sigcrb}). The data log summary is given
in Table~\ref{T1}.

\begin{figure*}[tb]
{\bf a) Radial velocities vs. phase\hspace{60mm} b) Radial
velocities vs. time}

\includegraphics[angle=0,width=8.5cm]{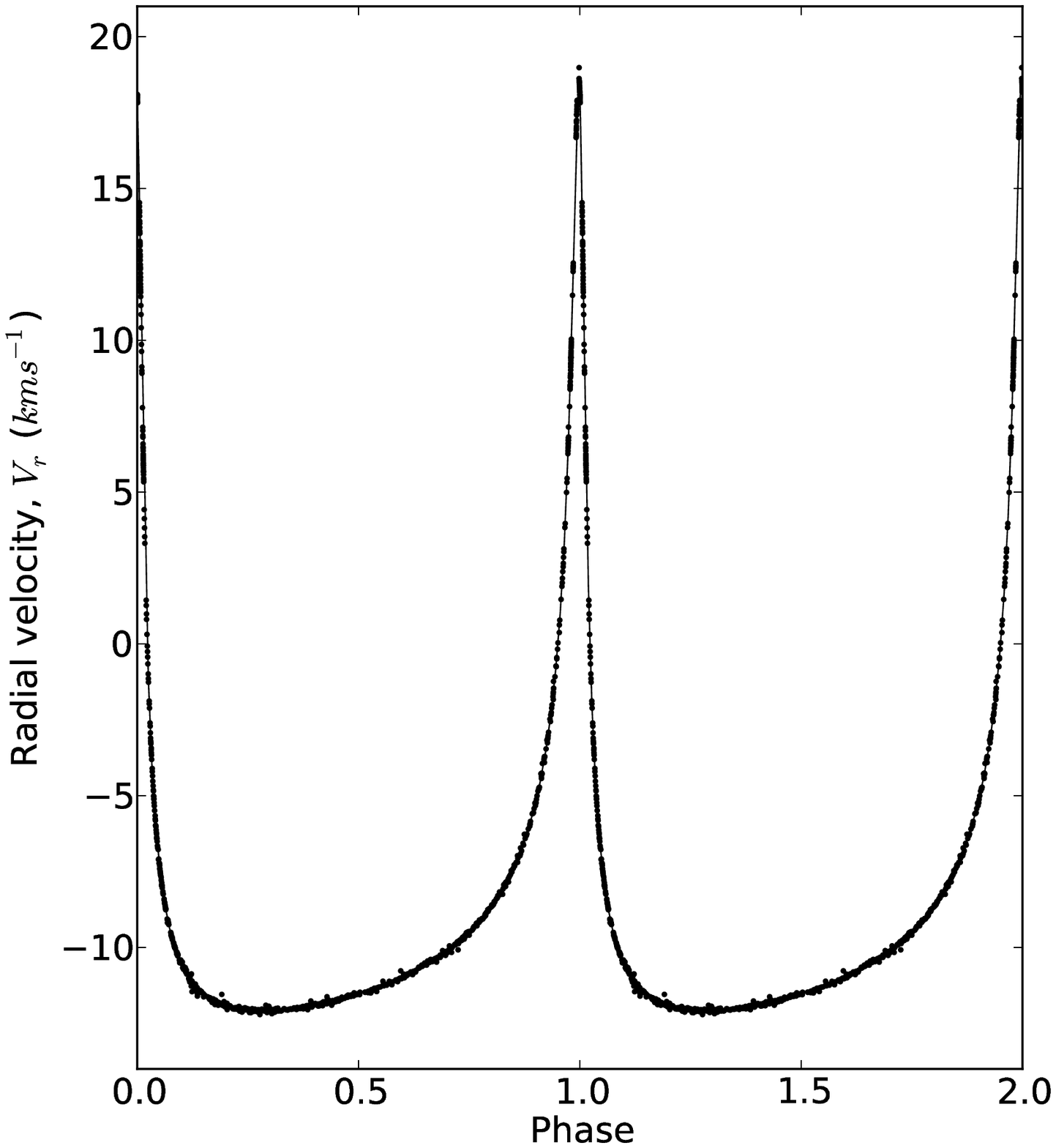}\hspace{5mm}
\includegraphics[angle=0,width=8.5cm]{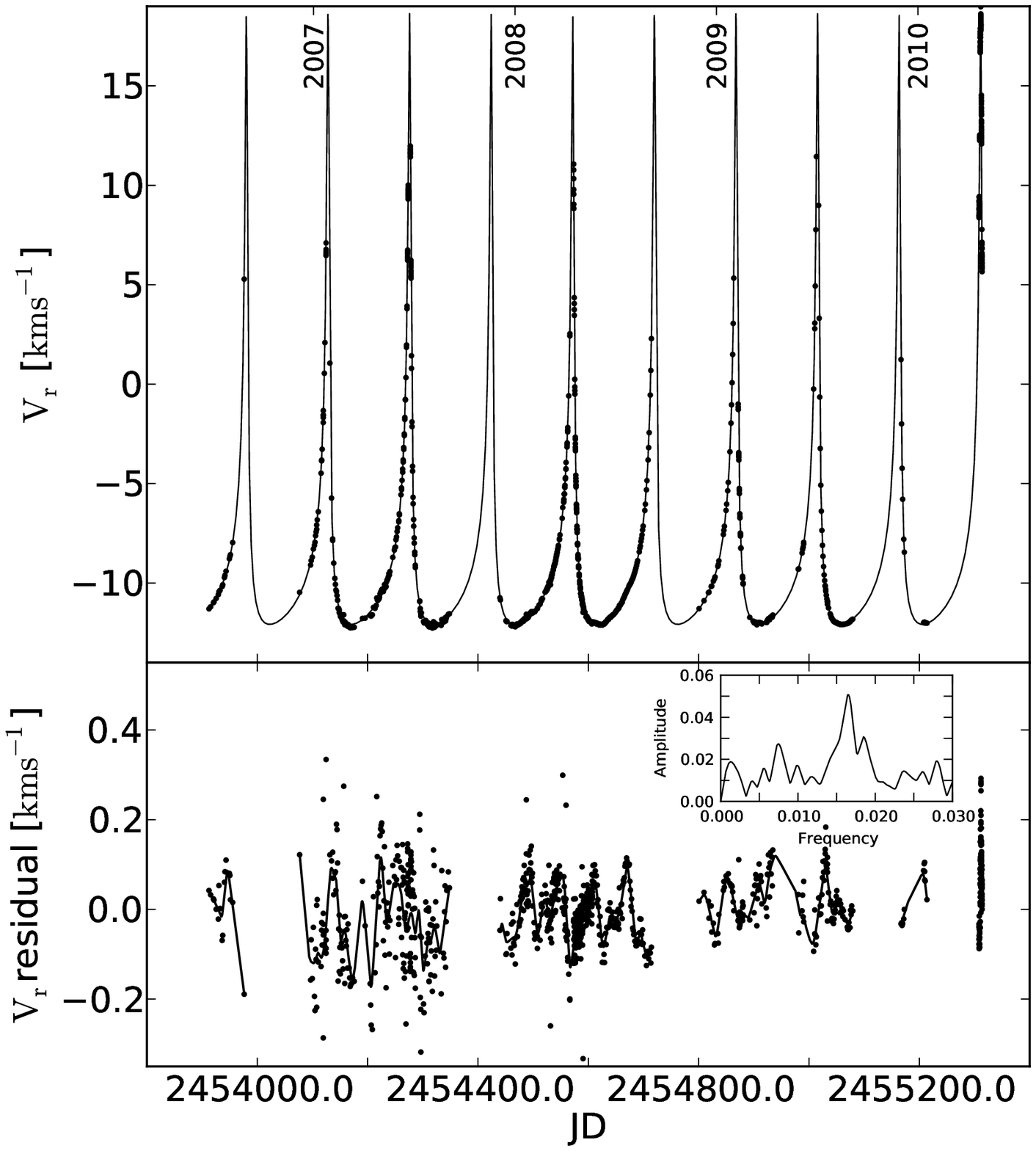}
\caption[ ]{Radial velocities of HD\,123351. {\bf a.} Phased
STELLA radial velocities with the orbital period of 147.8919~d.
Dots are the observations and the line is the orbital fit to the
spot-corrected elements in Table~\ref{T2}. {\bf b.} Top panel: The
four years of monitoring. Such data sampling is made possible with
robotic telescopes. The line is again the orbital solution from
Table~\ref{T2} (column ``Spot corrected''). Lower panel: Residuals
after removal of the orbital fit. The modulation is due to stellar
rotation and the insert shows a DFT periodogram with the strongest
peak at 61.0$\pm$1.1~d, in agreement with the photometric period.
The full amplitude varied over the years between 400~\ms\ in 2007
and 100~\ms\ in 2009, indicating the variable asymmetry of the
spot distribution. The line is a Blackman-function fit to the
data.
 \label{F2}}
\end{figure*}

\section{Stellar parameters of HD~123351}\label{S3}

\subsection{Orbital elements}\label{S3o}

Radial velocities of the STELLA spectra were determined from an
order-by-order cross correlation with a synthetic G8 template
computed from a Kurucz (\cite{kur}) atmosphere. Only the central
19 echelle orders are used, providing a wavelength range of
$\approx$150nm. The coud\'e spectra from KPNO and CFHT were
analyzed similarly but using the IRAF cross-correlation program
{\tt fxcor} together with various reference-star spectra. All
velocities are corrected for the solar-system's barycenter (the
full data are given in machine readable format in the electronic
attachment). The gravitational redshift is expected to be 133~\ms\
according to the mass and radius in Table~\ref{T3} but was not
corrected for.

For the orbital solution with non-zero eccentricity we use the
prescription from Danby \& Burkardt (\cite{dan:bur}) to calculate
the eccentric anomaly. The elements are solved for by using the
general least-square fitting algorithm {\em MPFIT} (Markwardt
\cite{mpfit}). More details about the orbit computation are
discussed in Weber \& Strassmeier (\cite{capella}) and we refer
the reader to this paper. Owing to unclear zero-point differences
between the three data sets, and the different set ups, we include
the earlier velocities from KPNO and CFHT only for the period
determination. An SB1 solution of all velocities, but with zero
weight given to the KPNO and CFHT data, resulted in an
eccentricity of 0.809$\pm$0.001 and a period of 147.890~d. We note
that all previous KPNO and CFHT velocities fell into the phase
range 0.1--0.8 where almost no velocity variations because of the
high eccentricity. From these spectra alone, one could conclude
that the star has a constant radial velocity, as actually done by
Strassmeier et al.~(\cite{ss}). For the initial orbital solution,
a total of 955 STELLA radial velocities were used. The standard
error in an observation of unit weight was 73~\ms, which is the
best so far achieved for an orbit of a chromospherically active
star (e.g. Fekel et al.~\cite{velo}, Griffin~\cite{griff09},
Torres et al. \cite{torr}). However, this rms includes the
modulation due to the asynchronously rotating spotted surface.
Table~\ref{T2} (left column) lists the orbital elements from the
unaltered initial data. In Sect.~\ref{S3rot}, we refine the orbit
by removing the starspot jitter from the radial velocity data and
then achieve a standard error of 47 \ms\ close to the measuring
error of approximately 30~\ms. Fig.~\ref{F2}a plots the
spot-corrected radial velocities along with the orbital fit,
Fig.~\ref{F2}b (top panel) shows the unaltered original data. The
difference cannot be seen by eye.

We note that the low mass function, together with our inability to
see a secondary star in our spectra, suggests that the orbit has a
low inclinationwith respect to the sky. If we assume at least a
2\fm5 brightness difference between the secondary and the primary
at red and blue wavelengths, the secondary cannot be of earlier
spectral type than G2(V). Adopting its mass as
$m_2$=1.0~M$_\odot$, the observed $f(m)$=0.0113 provides a lower
limit to the inclination of almost 22\degr . A more likely
secondary would be a red dwarf of mass $m_2\approx0.5$~M$_\odot$,
which suggests a more likely inclination of $\approx40$\degr\ and
a minimum separation of $\approx 2\,10^7$~km ($\approx$5 stellar
radii) during periastron.

\begin{table}[!tbh]
\begin{flushleft}
\caption{Spectroscopic orbital elements.}\label{T2}
\begin{tabular}{lll}
\hline \noalign{\smallskip}
Parameter & Original & Spot corrected\\
\noalign{\smallskip} \hline \noalign{\smallskip}
$P_{\rm orb}$ (days)        & 147.890$\pm$0.001 & 147.8919$\pm$0.0003\\
$T_{\rm periastron}$ (HJD245+) & 3980.616$\pm$0.004 & 3980.6098$\pm$0.0021\\
$\gamma$ (km~s$^{-1}$)      & $-$8.703$\pm$0.003 & $-$8.7057$\pm$0.0017\\
$K_{1}$ (km~s$^{-1}$)       & 15.35 $\pm$0.03 & 15.3518$\pm$0.0091\\
$e$                         & 0.809$\pm$0.001 & 0.8086$\pm$0.0001\\
$\omega$                    & 15.34 $\pm$0.01 & 15.326$\pm$0.018\\
$a_1\sin i$ (10$^6$ km)     & 18.35 $\pm$0.03 & 18.368$\pm$0.012\\
$f(m)$ (M$_{\sun}$)         & 0.01128$\pm$0.00010 & 0.011317$\pm$0.000023\\
fit rms (\ms )              & 73 &  47 \\
\hline
\end{tabular}

\vspace{1mm}Note. The columns are for the unaltered data
(``Original'') and for starspot-jitter removed data (``Spot
corrected'').
\end{flushleft}
\end{table}

\subsection{Rotational parameters}\label{S3rot}

The photometric data have high amplitude, both short-term and
long-term, variations. The short-term variations are clearly due
to rotational modulation, while the long-term variation is likely
due to some sort of multi-periodic spot cycle.

\begin{figure}[tb]
{\bf a) DFT periodogram}\\
\includegraphics[angle=0,width=7.9cm]{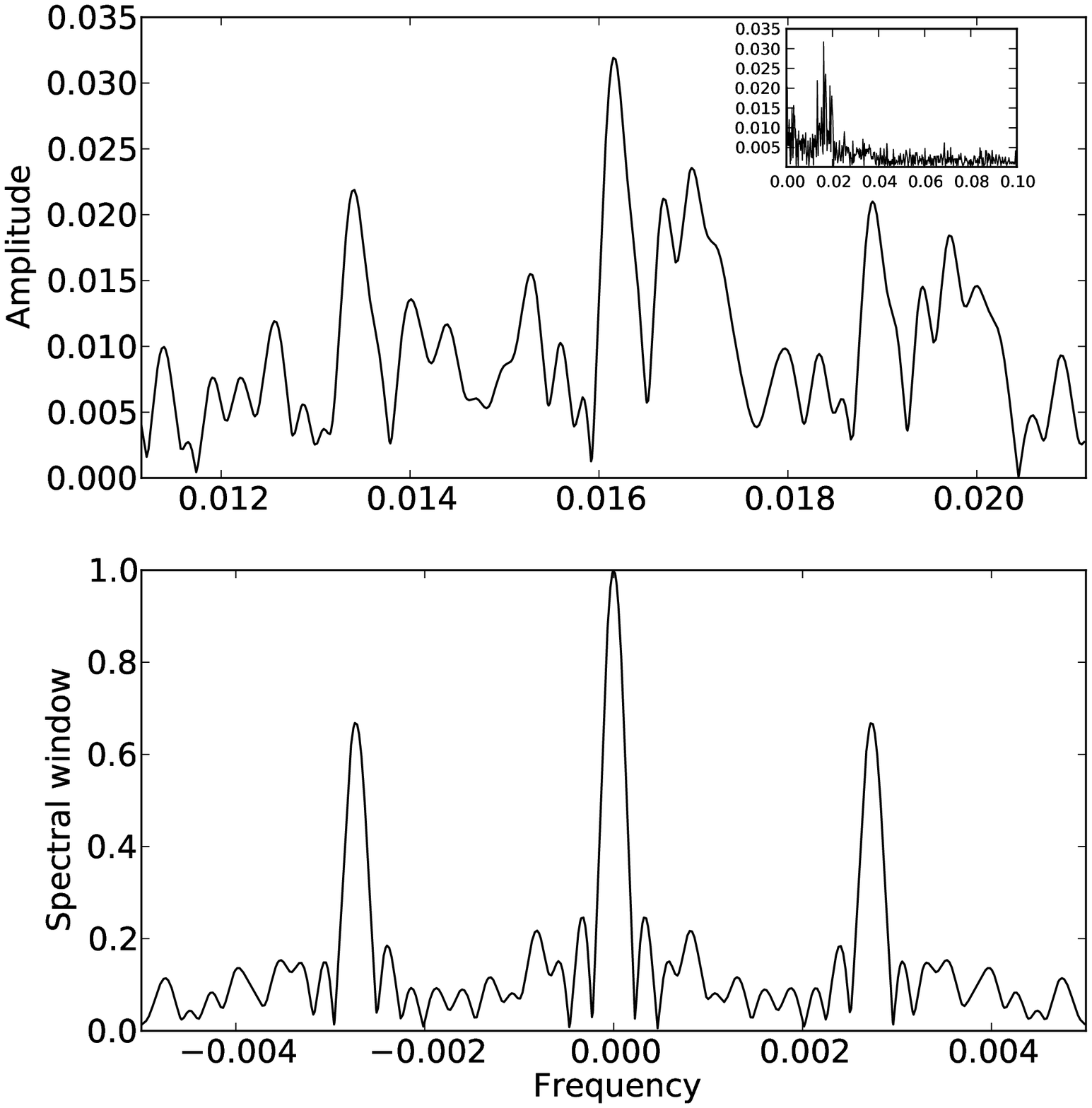}\\
{\bf b) STFT time-frequency analysis}\\
\vspace{1mm}
\includegraphics[angle=0,width=8.2cm,clip]{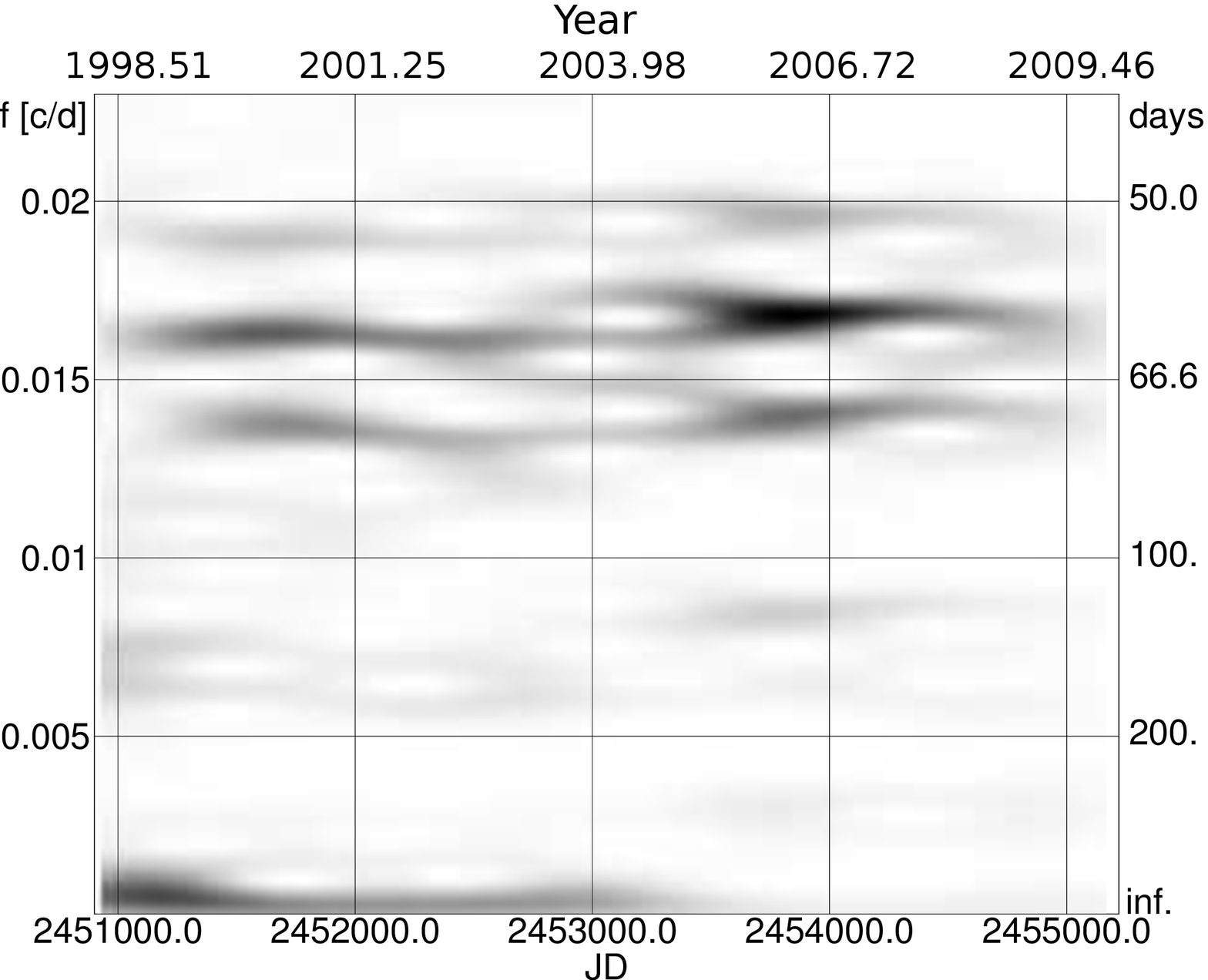}
\caption{Periodograms from HD\,123351 photometry. {\bf a.}
Discrete Fourier transform (DFT) periodogram from the pre-whitened
$V$-band data (top panel). The twelve-year average period is
61.9~d (0.01616 c/d). The bottom panel is the corresponding window
function. The insert in the top panel shows the entire search
range. {\bf b.} Two-dimensional short-term Fourier transform
(STFT) analysis. The power distribution in the time-frequency
plane is shown as a gray scale (black means highest power, white
no power). It shows that the photometric period reached values
between a maximum of 62.0~d and a minimum of 57.3~d. Around
2,453,000 (2002--2005), a bimodal distribution was recovered. Note
that the excess power in the lower left corner is due to the
length of the data vector and is artificial.
 \label{F3}}
\end{figure}

A photometric period of $\approx$60~d is obtained with a very high
level of confidence, which we interpret to be close to the true
stellar rotation period. This value is based on a one-dimensional
discrete Fourier transform (DFT) analysis of the $VI_C$-band data.
The best-fit twelve-year average period is 61.9$\pm$0.4~d from a
Gaussian fit to the $V$ and $I_C$-band DFT peaks (the results from
$V$ data are shown in Fig.~\ref{F3}a). However, other significant
peaks show up at slightly higher frequencies even in the
pre-whitened data. Repeating the analysis for three data subsets
by splitting the entire time series into three parts of similar
length, we get periods of 61.4~d for the first (1998--2001), 62.0
and 57.3~d for the second (2002--2005), and 59.2~d for the third
(2006-2010) part of the data set, identifying the variable and
sometimes even two significant periods. The typical error is
0.7~d. For spotted stars, this is a sign of surface differential
rotation, i.e., spots modulate the light with different periods
when they appear at different latitudes on the stellar surface. It
also means that we cannot simply interpret the photometric period
to be the rotational period of the star. Seasonal changes occur
when spots disappear after a certain (life)time but others appear,
possibly cyclically as on the Sun, and possibly at different
latitudes. If either the cycle is short enough or the database of
sufficiently long duration and the sampling dense enough, it
allows us in principle to determine the amount of differential
rotation (e.g. Ol\'ah et al.~\cite{ola:jur}). Its sign, however,
could only be determined from simultaneous Doppler imaging.

To confirm the above-mentioned period splitting, and pinpoint its
onset and duration, we performed a two-dimensional time-frequency
analysis with the program package Time-Frequency Analysis (TiFrAn)
(Koll\'ath \& Csubry \cite{tifran}). The time-frequency analysis
is conventionally applied to continuous datasets, whereas our
observations are gapped. Thus, the yearly interruptions are filled
with straight lines, which do not alter our main results but may
produce spurious features. More details about the procedure can be
found in Koll\'ath \& Ol\'ah (\cite{ko-ol}). We chose the
short-term Fourier transform (STFT) but also compared the results
with those obtained using the Choi-Williams and the pseudo-Wigner
kernel. The resulting patterns of periods are identical to those
of the above DFT analysis to a very high degree. Fig.~\ref{F3}b
shows the frequency distribution over the 12 years of data from
the STFT analysis. The continuous variability, and the splitting
of the rotational period, is clearly visible. The splitting
started in 2003 and lasted until 2005. The 2004 observing season
is the one where the light curve shape became double-humped and
the $V$ and $I_C$ amplitudes reached their smallest value in our
entire time series. At the same time, the photometry appeared
bluest, indicating, together with the low amplitudes, that the
spot coverage had reached a minimum.

The coadded $R=120,000$ CFHT spectrum with S/N=500 in the
wavelength range 638--646\,nm was measured for spectral-line
broadening. We followed the recipe of Fekel (\cite{fcf}) and found
a best-fit $v\sin i$ value of 1.8$\pm$0.7~\kms , adopting a
typical radial-tangential macroturbulence for a K subgiant/giant
of 3~\kms . We note that such a low $v\sin i$ cannot be resolved
by our spectra and thus formally constitutes only an upper limit.
Together with $P_{\rm phtm}\approx P_{\rm rot}\approx 60$~d and
the assumption that the rotational axis is perpendicular to the
orbital plane, we infer that the minimum stellar radius is $R\sin
i\approx$2.1$\pm$0.8~R$_\odot$.

\subsection{Rotation induced radial-velocity variations}

Today's high-precision radial-velocity surveys of stars with
exoplanets may record the rotation of the host star if it has dark
spots just large enough to mimic a ``transit'' (Saar \& Donahue
\cite{saa:don}). Such rotational modulation was first detected in
active pre-main sequence stars with large cool spots (e.g. Stelzer
et al. \cite{stel:v410}) but also in an active main-sequence star
(Huber et al. \cite{huber}). Amplitudes of 2~\kms\ were detected
for the extremely active T~Tauri star LkCa\,19 (Huerta et al.
\cite{huerta}), while a flare on the M-dwarf CN~Leo introduced
radial velocity jitter of several tens of \ms\ (Reiners
\cite{reiners}). We found radial velocity variations with an
amplitude of 270~\ms\ for the single, active giant 31~Com with a
period in agreement with the photometric period from MOST
satellite data (Strassmeier et al. \cite{31com}).

HD\,123351 exhibits radial-velocity modulations that are even
visible by eye in the plot of the residuals in Fig.~\ref{F2}b
after the removal of the orbital variation. The peak-to-peak
amplitude varied across the range 300--400~\ms\ in 2007 and was
100~\ms\ in early 2009. The simulations of Hatzes (\cite{hat})
suggested that spots induce a radial-velocity amplitude of at
least hundred m/s for even low-activity stars. His relationship of
velocity amplitude $\delta v$ in \ms\ with spot coverage $A$ in
per cent and line broadening $v\sin i$ in \kms\ ($\delta v=(8.6\,
v\sin i - 1.6) A^{0.9}$) would suggest a variable spot coverage in
the range of $A\approx 9-42$~\% for HD\,123351. This is incorrect
by a factor of three when compared to the more detailed
simulations by Desort et al. (\cite{des}), who found $3-12$~\% ,
mostly because these authors took into account the spot
temperature. However, this difference is of no relevance here
because we only wish to make an order-of-magnitude estimate.
Makarov et al. (\cite{mak:bei}) extended these analytical
relations to astrometry. We performed a DFT analysis of the
radial-velocity residuals shown in Fig.~\ref{F2}b and found the
best-fit period to be 61.0$\pm$1.1 days, in agreement with the
photometric period.

The rotation of the spotted stellar surface introduces systematic
jitter into the radial velocity curve. However, this jitter is
unevenly distributed across one orbital revolution because of the
asynchronous stellar rotation. We note that the residual
radial-velocity curve appeared even quintuple humped in 2008, just
as the light curve appeared double humped at the same time. We
remove the starspot jitter from the original data and then repeat
the determination of the orbital elements from Sect.~\ref{S3o}.
This is achieved by convolving the residuals in Fig.~\ref{F2}b
with a smoothing function (Blackman function) and then subtracting
it from the radial-velocity data. The largest values that were
added to or removed from the data were $+100$~\ms\ and $-200$~\ms
, respectively. The revised orbital elements are given in the
column ``Spot corrected'' in Table~\ref{T2}.

\begin{table}[tb]
\caption{Summary of astrophysical data of HD~123351. \label{T3}}
\begin{flushleft}
 \begin{tabular}{ll}
  \hline\noalign{\smallskip}
  Parameter                & Value    \\
  \noalign{\smallskip}\hline\noalign{\smallskip}
  Maximum $V$ magnitude    & 7\fm235 \\
  Spectral type            & K0 IV-III \\
  Hipparcos distance       & 97.8$^{+6.9}_{-6.0}$~pc \\
  $M_{\rm V}$              & +2\fm20$\pm$0.14 \\
  $A_V$                    & 0\fm08 \\
  $T_{\rm eff}$            & 4780$\pm$70 K   \\
  $\log g$                 & 3.25$\pm$0.30 \\
  $v\sin i$                & 1.8$\pm$0.7~\kms    \\
  Rotation period          & 58.32$\pm$0.01 d \\
  Inclination              & $\approx$40\degr   \\
  Radius                   & 5.74$^{+0.69}_{-0.62}$ R$_\odot$  \\
  Luminosity               & 15.4$^{+2.8}_{-2.3}$~L$_\odot$ \\
  Mass                     & 1.2$\pm$0.1 M$_\odot$  \\
  Age                      & $\approx$6--7 Gyr  \\
  $[$Fe/H$]$               & $0.00\pm0.08$ \\
  Lithium abundance        & 1.70$\pm$0.05\\
  Average turn-over time   & 40~d \\
  Microturbulence          & 1.15 \kms \\
  Macroturbulence          & 3.9 \kms \\
  Average magnetic flux    & 542$\pm$72 G \\
  \noalign{\smallskip}\hline
 \end{tabular}
\end{flushleft}
\end{table}

The effect of the applied procedure on the orbital elements is
also demonstrated in Table~\ref{T2}. We note that the elements did
not change much, i.e. the corrected solution is largely within the
errors of the uncorrected solution. Typically, the error per
element became smaller by a factor 2-3. This is what is expected
if the jitter due to rotational modulation spreads equally, or
nearly equally, around a mean radial velocity. For our long-term
observations, this is indeed the case because the phased orbital
radial-velocity curve is sampled in total by 25 stellar rotations
(for each orbital revolution we have 2.5 stellar rotations).
Because the starspot distribution changes with time, we do not
just correct with a harmonic sinusoidal of period $P_{\rm rot}$
and fixed amplitude but with the actual residuals of the
rotationally modulated radial velocity.

\subsection{Luminosity, radius, metallicity, mass, and age}

\begin{figure}
\includegraphics[angle=0,width=86mm,clip]{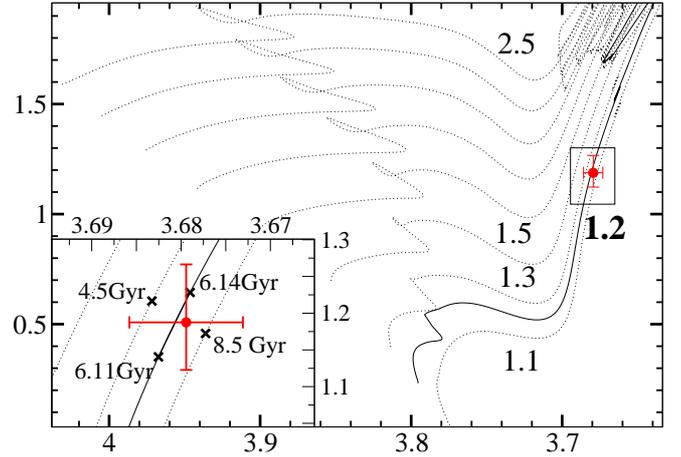}
\caption{The position of HD~123351 (dot) in the theoretical H-R
diagram ($\log L/L_\odot$ vs. $\log T$ in K). Shown are
evolutionary tracks for masses of 1.1--2.5~M$_\odot$, solar
metallicity, and core overshooting (dotted lines; Pietrinferni et
al. \cite{piet}). The best-fit model track (full line) suggests a
mass for HD~123351 of 1.2$\pm$0.1~M$_\odot$. The insert is a
zoomed image of the box around the observed location indicating
the nominal ages of the 1.2-M$_\odot$ track (again the full line)
and a 1.1-M$_\odot$, 1.3-M$_\odot$, and 1.5-M$_\odot$ track
(dotted lines), respectively. This suggests a most-likely age of
HD~123351 of 6--7~Gyr.}\label{F4}
\end{figure}

Fig.~\ref{F4} compares the position of HD~123351 in the H-R
diagram with theoretical evolutionary tracks. We note that the
revised \emph{Hipparcos} parallax of 10.22$\pm$0.67~mas (van
Leeuwen \cite{nhip}) corresponds now to a distance of
$97.8^{+6.9}_{-6.0}$~pc, which is different by +1.7~pc from the
original data reduction. At this distance, it is only moderately
affected by interstellar extinction. We adopted the mean
extinction value from Henry et al. (\cite{hetal00}) for the
absolute magnitude calculation ($A_{\rm V}$=0.8 mag~kpc$^{-1}$ and
$E(B-V)=A_{\rm V}/3.3$). The brightest recorded $V$ magnitude of
7\fm235 converts into an absolute visual magnitude of $M_{\rm V}$
= $+2\fm20\pm0.14$. Its dereddened $B-V$ based on the Tycho color
(1\fm029$\pm$0.01) is 1\fm00. This can be converted to a $T_{\rm
eff}$ of $\approx$4840~K with the transformation of Flower
(\cite{f96}). The bolometric magnitude of HD~123351 is +1\fm78
obtained with a solar bolometric magnitude of +4\fm75 and a
bolometric correction ($B.C.$) of $-0.42$ (Flower \cite{f96}). The
$\pm$70~K uncertainty in $T_{\rm eff}$ (see next paragraph)
corresponds to an uncertainty in the $B.C.$ of $\pm$0.04 and the
star's luminosity becomes 15.4$^{+2.8}_{-2.3}$~L$_\odot$. These
errors are still driven by the error in the parallax rather than
in the effective temperature ($\pm$70~K). The Stefan-Boltzmann law
gives a stellar radius of 5.74$^{+0.69}_{-0.62}$~R$_\odot$, which
is appropriate for a K0 subgiant/giant. The minimum radius
inferred from $v\sin i$ and $P_{\rm rot}$ in Sect.~\ref{S3rot} of
2.1$\pm$0.8~R$_\odot$ suggests a minimum inclination of as low as
$i\approx 23\degr$.

Our SES spectra are used to refine the effective temperature,
gravity, and metallicity. We employ the spectrum-synthesis
technique and the numerical tools described in Allende-Prieto et
al.~(\cite{all04}) with a grid of synthetic ATLAS-9 spectra
tailored to the stellar parameters of HD\,123351. Five spectral
orders around 600~nm were chosen and the result per spectral order
combined on the basis of a weighted least-squares minimization.
The average and the standard deviations then constitute our final
values and their internal precisions. We found $T_{\rm
eff}=4780\pm70$~K, $\log g$=3.25$\pm$0.30, and
[Fe/H]=$0.00\pm0.08$ with a prefixed value for the microturbulence
of 2~\kms\ and a total line broadening of 3.80~\kms.

These values are compared with theoretical evolutionary tracks in
Fig.~\ref{F4}. We selected post main-sequence tracks from the grid
of Pietrinferni et al. (\cite{piet}). All tracks are for solar
metallicity and core overshooting. The HD~123351 position suggests
a most likely mass of 1.2$\pm$0.1~M$_{\odot}$. We note that the
models without core overshooting would increase the mass by about
0.1~M$_{\odot}$ and decrease the age by $\approx$1~Gyr. The star
appears on the ascent of the red-giant branch (RGB) at an age of
approximately 6--7~Gyr. The $^{12}$C$^{14}$N lines around
800.35~nm are clearly detectable with a combined equivalent width
of 96~m\AA . However, the equivalent width of the $^{13}$C$^{14}$N
line at 800.46~nm is at $\approx$2.5~m\AA\ very weak and probably
below our detection limit but provides a lower limit to
$^{12}$C/$^{13}$C of $\approx$40, in agreement with the
expectation that small $^{12}$C/$^{13}$C ratios (around 10--15)
are only found in giants that have evolved up to or beyond the RGB
(e.g. Allende Prieto et al.~\cite{all}). At this point we recall
the findings of radius and temperature inconsistencies for active
low-mass eclipsing binaries with model computations (e.g. Morales
et al. \cite{mor2009}, \cite{mor2010}). Oversized stars with a
temperature deficit caused by chromospheric activity could alter
the true stellar position in the H-R diagram, thus affect the mass
and age estimate.


\subsection{Lithium abundance}

\begin{figure}
\includegraphics[angle=-90,width=8.6cm]{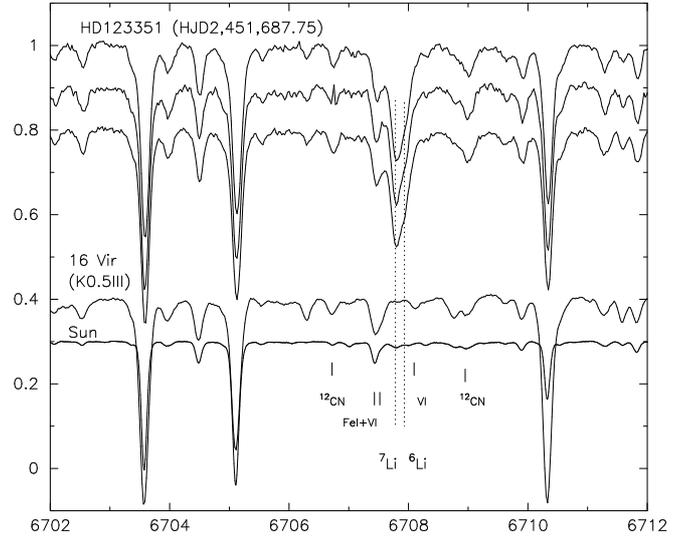}
\caption{Lithium 670.8-nm region in the spectrum of HD\,123351.
Shown are three consecutive $R=120,000$ spectra obtained at CFHT
in May 2000 at orbital phase 0\fp496 along with a reference
spectrum of the inactive K0.5III star 16~Vir. All spectra are
shifted in intensity for a clearer display. A very high-resolution
spectrum of the Sun from the National Solar Observatory solar
atlas (Kurucz et al. \cite{nso}) is shown for comparison.
HD\,123351 shows a significant lithium line with a total
equivalent width of 90~m\AA. } \label{F5}
\end{figure}

We employ the averaged $R$=120,000 CFHT spectrum for the
measurement of the Li equivalent width (Fig.~\ref{F5}). This
spectrum has a peak S/N of almost 500:1. A Gaussian fit gives an
equivalent width of the combined 670.78~nm line ($^6$Li+$^7$Li) of
90$\pm$3~m\AA \ (the error is an estimate inferred from the spread
in the values from the three individual spectra and the quality of
the Gaussian fit). Subtraction of a reference spectrum with
presumably no lithium (16~Vir in Fig.~\ref{F5}) removes a
contribution of 16~m\AA\ from the nearby Fe\,{\sc i}+V\,{\sc i}
blend. The remaining 74~m\AA\ converts into an abundance of $\log
n$(Li)=1.70 ($\log n$(H)$\equiv$12) with $T_{\rm eff}=4780$~K and
$\log g = 3.0$ from the NLTE tables in Pavlenko \& Magazz\'u
(\cite{pav:mag}). Its uncertainties produced by the measuring
error and the error of $\pm$70~K in $T_{\rm eff}$ and $\pm$0.3 in
$\log g$, combine to $\approx$0.05~dex.

The very low rotational broadening of $v\sin i=1.8$~\kms \ enables
a detailed line-profile modeling of the $^6$Li/\,$^7$Li isotope
ratio. As this is a complex synthesis based on a
three-dimensional, radiative hydrodynamics simulation, we refer to
a forthcoming paper by Steffen et al. (\cite{steffen}). Not much
is known about this ratio in magnetically active stars (see, e.g.,
Balachandran et al.~\cite{bal:fek}) but it plays an important role
in constraining or even excluding some mixing mechanisms and
determining the likelihood of extra lithium production. HD\,123351
shows a moderately strong but broader than expected Li\,{\sc i}
line at 670.8~nm (see Fig.~\ref{F5}). The line strength is weaker
than to the nearby Fe\,{\sc i} lines at 671.03~nm and 670.51~nm,
i.e. 0.76 compared to 0.68 and 0.67, respectively, but the
equivalent width is larger by more than 20\%. A weak excited Li
feature at 610.36~nm is also present but strongly blended with the
nearby double Fe\,{\sc i} line and appears rather uncertain for a
quantitative analysis. We measure an equivalent width of
8$\pm$4~m\AA\ from a double Gaussian fit, close to the detection
limit of, say, 3--5~m\AA. Therefore, we refrain from modelling
this line.

\subsection{Rossby number and dynamo activity}

\begin{figure}[tb]
\includegraphics[angle=0,width=5cm]{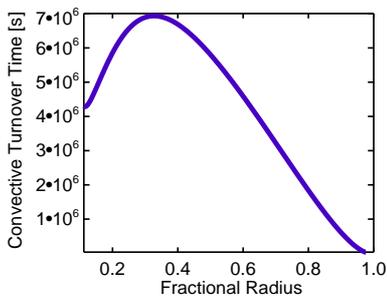}
\caption[ ]{Convective turn-over times, $\tau_c$, in seconds as a
function of depth in the convection zone of HD~123351. Note that the
turn-around of the curve below a fractional radius of $\approx$0.35
strongly depends on the adopted stellar model and could be an
artefact. $\tau_c$ spans between a maximum of 80~d near the bottom
of the convection zone and 10~d near the surface.
 \label{F6}}
\end{figure}

The Rossby number, $Ro=P_{\rm rot}/\tau_c$, is widely used as an
indicator of the existence and efficiency of a dynamo process.
While the rotation period can be observed very accurately, the
convective turn-over time, $\tau_c$, remains an elusive parameter.
It describes the ratio of a characteristic length of convection to
a characteristic velocity, $\ell/u$, and is either determined
theoretically (e.g. Kim \& Demarque \cite{kim:dem}) or
semi-empirically (e.g. Pizzolato et al.~\cite{pizz}). We determine
$\tau_c$ with the differential-rotation code of K\"uker \& Stix
(\cite{kue:sti}) as a function of depth based on a description of
the pressure scale height, $H_p$, and the classical mixing length,
$\alpha$. The characteristic convective length is then defined as
$\ell = \alpha\ H_p$, with $\alpha = 2$. In the case of a fast
rotator, the convective heat flux is no longer aligned with the
temperature gradient and tilts towards the rotational axis. It
then creates a horizontal heat flux from the equator to the poles
(meridionally circulated) that, together with the Reynolds stress,
maintains differential rotation. Fig.~\ref{F6} shows the turn-over
time for HD~123351 plotted versus the fractional stellar radius.
The depth of the convection zone is predicted from the
MESA\footnote{http://mesa.sourceforge.ne} code (Paxton et al.
\cite{mesa}) to be 90\%\ of the stellar radius. The average
turn-over time is around 40~d. It would give a Rossby number of
$\approx$1.4 and, according to the dynamo criterium of Durney \&
Latour (\cite{dur:lat}), be insufficient to operate an effective
dynamo, in contradiction with the observations. If we determine
the Rossby number using the convective turnover time at the base
of the stellar convection zone, i.e. $Ro\approx$0.73 with
$\tau_c\approx 80$~d, it agrees with the dynamo criterium. This
may indicate that the dynamo of HD~123351 is indeed located at the
base of the convection zone, as in the Sun.

\subsection{Chromospheric activity}

We measure an average equivalent width of the He\,{\sc i} D3-line
of 39~m\AA\ with an estimated error of 5~m\AA , which is
comparable to the measurements for most active G and K dwarfs in
the sample of Saar et al. (\cite{saar-he1}). The detection of the
He\,{\sc i} D3-line at 587.56~nm indicates the presence of
non-thermal heating processes on HD\,123351, since the line cannot
be excited at photospheric temperatures. It is thus a comparably
``pure'' indicator of chromospheric magnetic activity because of
the absence of photospheric contamination present in most other
optical activity diagnostics, e.g. Ca\,{\sc ii} H\&K and \Halpha .
However, the comparable weakness of the line makes it more likely
to be affected by a limited S/N ratio.

Besides He\,{\sc i} D3, both Ca\,{\sc ii} features, the H\&K
doublet, and the infrared triplet, appear as moderately-strong
central emission lines, while \Halpha\ appears as a filled-in
absorption line with an equivalent width of 850~m\AA . Absolute
Ca\,{\sc ii} H\&K emission-line fluxes for HD\,123351 are
determined with the method of Linsky et al. (\cite{lin:etal}) as
adapted and applied, e.g., in Strassmeier et al. (\cite{str:fek}).
It is based on the measurement of the relative flux in a 5.0-nm
bandpass centered at 395.0~nm, and the relative flux in the two
emission lines, defined by the blue and red emission-line
footpoints, respectively. Absolute fluxes are then determined
using the Barnes-Evans relation (Barnes \& Evans \cite{be}) and
the measured value of the star's Johnson $V-R$ color. Following
the recipes in Hall (\cite{hall}), the relative fluxes convert
into absolute chromospheric emission-line fluxes of, on average,
1.6$\times$10$^6$ erg\,cm$^{-2}$s$^{-1}$ for the sum of the
Ca\,{\sc ii} H and K lines, 1.4$\times$10$^6$
erg\,cm$^{-2}$s$^{-1}$ for \Halpha\ and 8$\times$10$^5$
erg\,cm$^{-2}$s$^{-1}$ for the sum of the Ca\,{\sc ii}
infrared-triplet lines. These are values typical of RS~CVn stars
and indicative of a magnetically active atmosphere.

\begin{table*}
\begin{flushleft}
\caption{Seasonal average spot parameters 1998--2009. }\label{T4}
\begin{tabular}{llllllllll}
\hline \noalign{\smallskip}
Year & mid HJD & $\ell_1$ & $\ell_2$ & $b_1$ & $b_2$ & $A_1$& $A_2$ & $T_{\rm spot}$ & $\sigma_{\rm fit}$ \\
      & & (\degr) & (\degr) & (\degr) & (\degr) & (\%) & (\%) & (K) & (mmag) \\
\noalign{\smallskip} \hline \noalign{\smallskip}
1998&2450943.815&
313.8$\pm$2.0 & 163.5$\pm$4.3& 56$\pm$4 & 55$\pm$9
& 7.45$\pm$0.08& 2.93$\pm$0.07& 3830$\pm$20 & 6.0\\
1999&2451264.194& 307.9$\pm$3.0 & 105.7$\pm$2.3& 52$\pm$6 & 53$\pm$4
& 4.07$\pm$0.06& 5.83$\pm$0.07& 3880$\pm$25 & 5.6\\
2000&2451620.425& 202.0$\pm$4.1 & 61.5$\pm$7.9& 55$\pm$8 &
20$\pm$22& 4.42$\pm$0.07& 1.91$\pm$0.17& 3250$\pm$135 & 12.4\\
2001&2452026.519& 257.6$\pm$2.3 & 52.1$\pm$4.4 & 61$\pm$5 &
20$\pm$10& 4.31$\pm$0.03& 2.19$\pm$0.09& 3210$\pm$75 & 9.3\\
2002&2452359.676& 238.5$\pm$2.7 & 52.1$\pm$1.7 & 23$\pm$8 & 55$\pm$4
& 1.88$\pm$0.04& 3.61$\pm$0.02& 3500$\pm$60 & 7.6\\
2003&2452702.954& 223.6$\pm$1.0 & 63.7$\pm$3.4 & 48$\pm$3 & 12$\pm$8
& 3.83$\pm$0.03& 1.40$\pm$0.05& 3420$\pm$60 & 5.0\\
2004&2453060.370& 249.2$\pm$2.1 & 75.2$\pm$4.4 & 41$\pm$5 & 26$\pm$9
& 3.23$\pm$0.03& 1.57$\pm$0.05& 3520$\pm$90 & 6.8\\
2005&2453444.012& 167.5$\pm$2.3 & 47.8$\pm$4.2 & 20$\pm$5 & 32$\pm$6
& 6.29$\pm$0.18& 3.36$\pm$0.12& 3850$\pm$80 & 7.6\\
2006&2453830.231& 183.6$\pm$1.7 & 43.7$\pm$5.4 & 36$\pm$4 &
18$\pm$10& 4.30$\pm$0.06& 1.51$\pm$0.09& 3525$\pm$100 & 6.1\\
2007&2454198.866& 184.4$\pm$2.3 & -37.1$\pm$3.0& 28$\pm$5 & 44$\pm$5
& 4.00$\pm$0.10& 3.05$\pm$0.07& 3600$\pm$70 & 6.8\\
2008&2454567.950& 204.8$\pm$1.5 & 22.5$\pm$4.1 & 26$\pm$9 &
66$\pm$14& 6.46$\pm$0.06& 2.86$\pm$0.03& 3970$\pm$70 & 5.5\\
2009&2454885.704& 245.4$\pm$7.5 & 30.7$\pm$7.7& 30$\pm$15&
42$\pm$17& 5.07$\pm$0.25& 4.72$\pm$0.19& 3870$\pm$180 & 20.5\\
\hline
\end{tabular}

\vspace{1mm}Note. Both spots have equal temperature. The errors come
from the summed residuals from the simultaneous light-curve fit in
$V$ and $I_C$. The stellar rotation period was fixed to 58.32~d.
\end{flushleft}
\end{table*}

\begin{figure*}[tb]
{\bf a) \hspace{6cm} b)}\\
\includegraphics[angle=0,width=6cm]{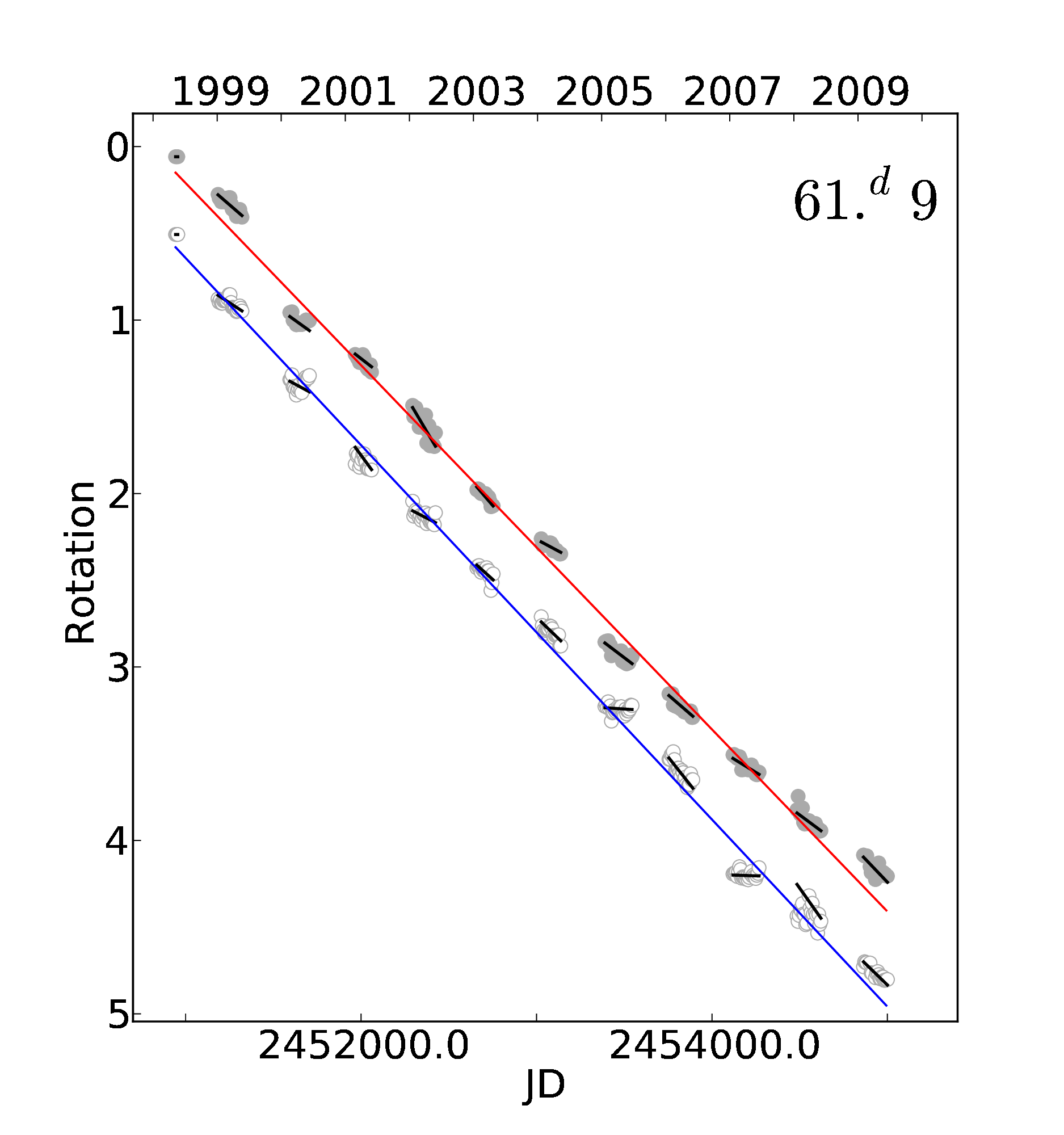}\hspace{3mm}
\includegraphics[angle=0,width=12cm]{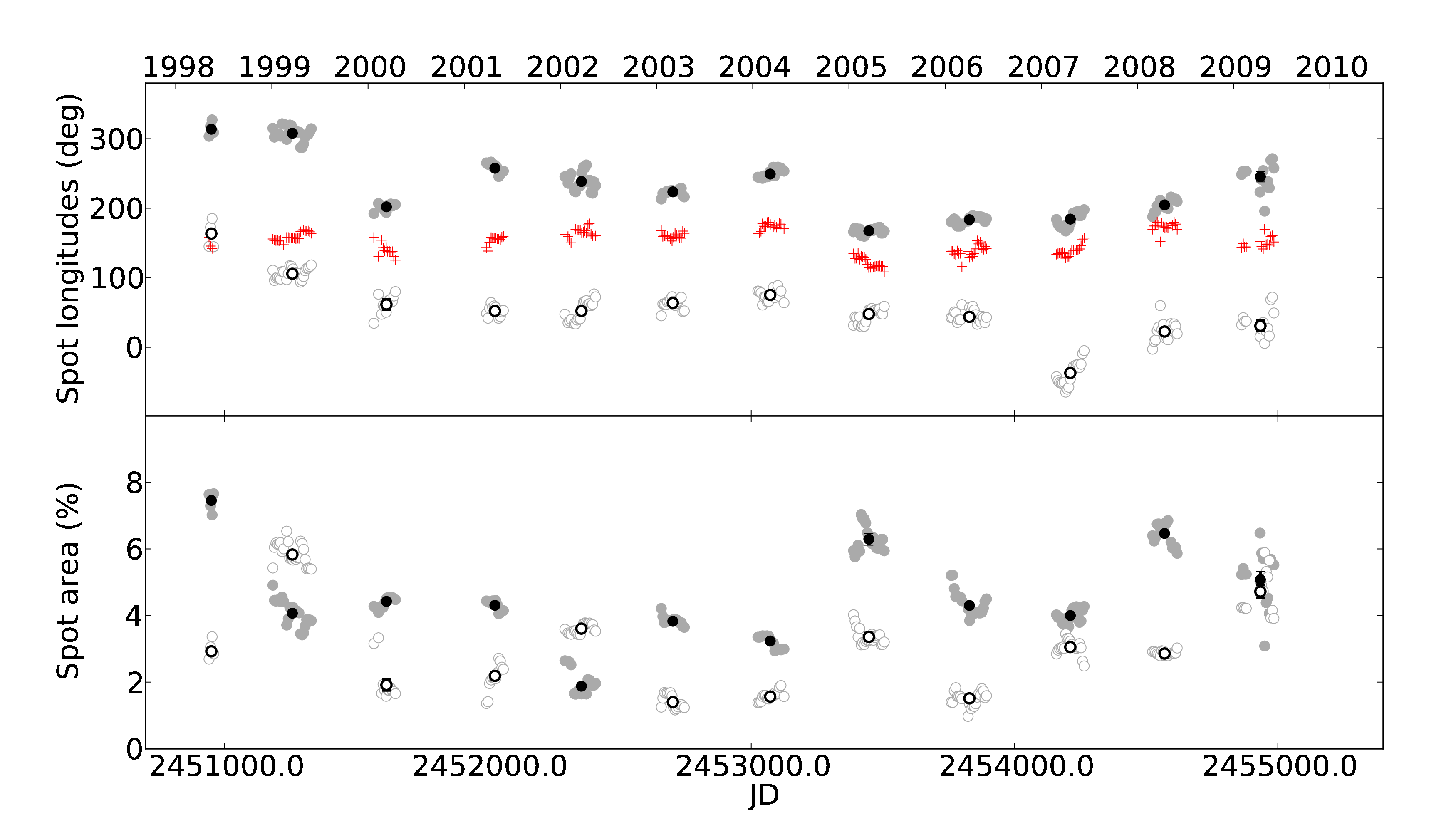}
\caption[ ]{Results from the photometric spot modeling. {\bf a}
The reconstructed spot longitudes expressed in stellar rotations
versus time. Filled gray points refer to spot~1, open circles to
spot~2. These values were obtained based on the initial
photometric period of 61.9~d and after shifting full seasons by an
integer rotation such that the resulting long-term migration is
brought into agreement with the seasonal migration (the latter is
indicated by the short lines). The two lines are linear regression
fits and their slopes indicate the true rotation period of
58.32~d. \ {\bf b} Top panel: spot longitudes versus time based on
the 58.32-d period. Filled gray points refer to spot~1, open
circles to spot~2 as in panel a. The pluses indicate the
longitudinal separation of the two spots and show a fairly
constant value of 150\degr$\pm$17\degr. Bottom panel: the
reconstructed fractional spot area versus time. In both panels,
seasonal average values are additionally plotted.
 \label{F7}}
\end{figure*}

\section{Time-series analysis}\label{S4}

\subsection{The location and evolution of photospheric
spots}\label{S41}

%
The spot longitude, which is the most stable and well-determined
parameter, arises directly from the light-curve shape and marks
the presumed spot concentration. The longitude difference between
the spots can depict systematic surface patterns. When the spots
are very close, then a single, elongated spotted region probably
causes the light variation that then appears quasi-sinusoidal. On
the other hand, double-humped light curves indicate spotted
regions on the two hemispheres of the star separated by nearly
180\degr . We perform a more detailed light-curve fitting with our
time-series spot-modelling code {\sc SpotModeL}. The strengths and
shortcomings of the code were described in detail in Rib\'arik et
al.~(\cite{sml}). The program reconstructs the position, size, and
temperature of up to three cool spots by minimizing the fit
residuals with the help of the Levenberg-Marquardt algorithm as
formulated in Press et al. (\cite{press}). Free parameters are the
spot radii, the temperature relative to the (fixed) photospheric
temperature, and their longitudes and latitudes. In the present
application, we solve for the $V$ and $I_C$ light curves
simultaneously. Figure~\ref{F1} illustrates the fits along with a
graphic presentation of the resulting two spot models.

\begin{table}
\begin{flushleft}
\caption{Summary of period determinations.}\label{T5}
\begin{tabular}{llll}
\hline \noalign{\smallskip}
Indicator & Period & Technique & FAP\\
          & (days) & & (\%) \\
\noalign{\smallskip} \hline \noalign{\smallskip}
$VI$-band photometry    & 61.9$\pm$0.4   & DFT   & \dots \\
Radial velocities       & 147.8919       & Orbit & \dots \\
Residual radial velocities & 61.0$\pm$1.1& DFT   & \dots \\
Spot migration          & 58.32$\pm$0.01 & LCM   & \dots \\
\Halpha\ line-core flux & 58.7$\pm$0.2   & DFT   & \dots \\
Ca\,{\sc ii} H\&K flux  & 63$\pm$5       & DFT   & \dots \\
Ca\,{\sc ii} H\&K flux  & $\approx$147   & DFT   & \dots \\
He\,{\sc i} D3          & 60.68$\pm$0.15 & Lomb  & 1\,10$^{-6}$\\
Magnetic flux           & 58.7$\pm$4.3   & Lomb  & 3\,10$^{-6}$\\
Magnetic flux           & 144.3$\pm$8.9  & Lomb  & 0.1 \\
\noalign{\smallskip}\hline
\end{tabular}

\vspace{1mm}DFT: discrete Fourier transform. LCM: light-curve
modeling. Lomb: Lomb-Scargle. FAP: false alarm probability.
\end{flushleft}
\end{table}

We first proceed with the determination of the spot temperatures
and the best-fit spot latitudes. For this initial step, we assume
the average photometric period (61.9~days) to be the rotational
period. Average spot temperatures are obtained by a simultaneous
two-spot fit of the seasonal $V$ and $I_C$ light curves. The
seasonal light curves usually span three stellar rotations with
significant changes in the light-curve shape in $V$ and in $I_C$.
Spot latitudes are kept as a free parameter in this initial step.
We note that these spot latitudes are strongly model dependent.
Here we assume two circular spots. The average rms of the fits to
$V-I_C$ implies spot temperatures precise to roughly $\pm$100~K,
with a span of 20--180~K for the various seasons. Any seasonal
change below this threshold is deemed unrealistic from our data.
The grand average spot temperature over the 12 years was 3620~K
(rms of 250~K), i.e. cooler than the unspotted photosphere by
1160~K. With the exception of years 2000 and 2009, all seasonal
rms values are below or around 100~K, i.e. 1.4 to 8 times smaller
than the 250~K from the grand average, thus considered real
differences. In a second step, we again applied {\sc SpotModeL}
but now to the $V$ and $I_C$ light curves in the time domain and
with the spot temperatures and spot latitudes kept fixed to the
corresponding average seasonal values listed in Table~\ref{T4}. In
that table, $\ell$ denotes the spot longitude, $b$ the spot
latitude (positive in the dominating hemisphere), $A$ the
fractional spot coverage in per cent of the total sphere, $T_{\rm
spot}$ the spot temperature, and $\sigma_{\rm fit}$ the squared
sum of the residuals in $V$ and $I$ in mmag. Labels 1 and 2 denote
spots~1 and 2, respectively.

\begin{figure*}[tb]
\includegraphics[angle=0,width=\textwidth]{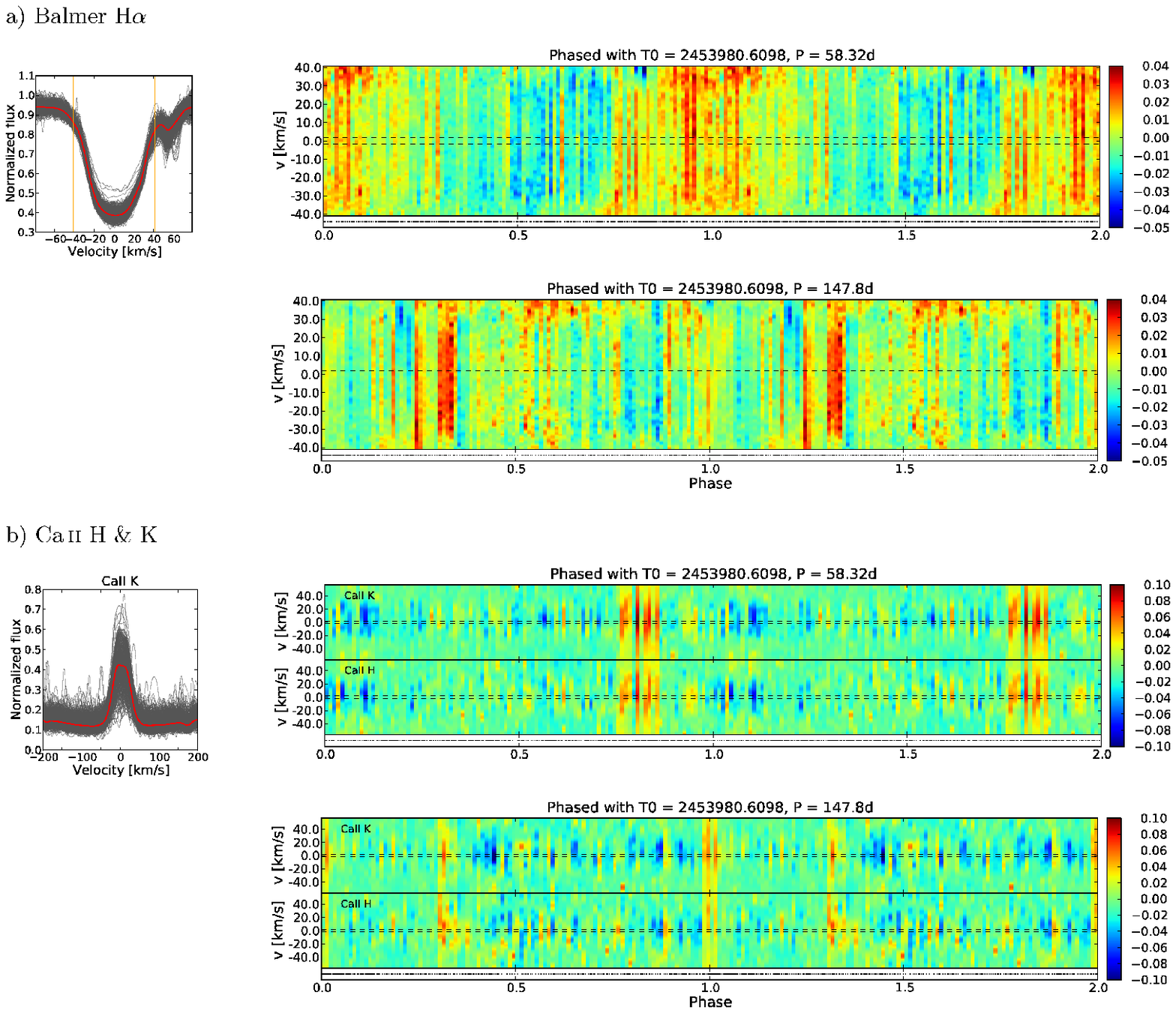}
\caption[ ]{Modulation of chromospheric activity tracers during
the years 2007--2010. The two top rows show \Halpha\ ({\bf a}),
the two bottom rows show Ca\,{\sc ii} H\&K ({\bf b}). The left
panels are overplots of all line profiles for 2007--2010 (in panel
b only the K-line is shown), while the bright lines indicate the
median profiles, respectively. The right hand panels are the
dynamic spectra phased with the photometric rotational period
(top) and the orbital period (bottom) and the zero points from
Table~\ref{T2}. The color scale indicates the relative
emission-line flux with respect to the median profile. The two
dashed, horizontal lines show $\pm v\sin i$, and the chain of
points in the very bottom of the graphs represents the phase
coverage (one dot is one spectrum). It appears that both activity
tracers display a clear modulation with the rotational period,
with a tendency to cluster around a time of periastron at phase
1.0, and maintained that behavior for the entire duration of our
observations.
 \label{F8}}
\end{figure*}

The next step included the revision of the rotational period. This
was done by determining the slope of the longitudinal spot
migration in the longitude versus (vs.) time diagram for each spot
and during each observing season. Then, the seasonally averaged
longitude of each spot, or spot group, was allowed to be shifted
by an integer rotation such that the resulting long-term migration
was brought into the best-possible agreement with the seasonal
migration (Fig.~\ref{F7}a). This was achieved for a slope of
$-0.38$~$\deg/d$ for spot~1 and $-0.39$ for spot~2, corresponding
to a new period of 58.32~days. As a comparison, the seasonal
average slopes were $-0.32$ and $-0.35$ for spots~1 and 2,
respectively. The slight difference between the two spots is due
to spot~1 appearing at a slightly longer longitude than expected
in 2000 and by spot~2 being at a markedly shorter longitude than
expected in 2007. We tend to ignore the difference and adopt the
average slope as the true rotation period of the star. We then
applied {\sc SpotModeL} again in the time domain but with the
rotation period fixed to 58.32 days. This led to the refined spot
parameters in Table~\ref{T4}. We emphasize that our spot
solutions, although precise and consistent, are not a detailed
accurate representation of the real stellar surface because of the
assumption of circular spots and the mathematically non-unique fit
(see Savanov \& Strassmeier \cite{sav:str}) but serve as a
qualitative surface map for further discussions.

Fig.~\ref{F7}b shows the migration of the spot longitudes and the
spot size together with the (longitudinal) separation of the two
spotted regions based on a period of 58.32 days. We note that in
all cases only two spots were needed to fit the light curves to
within their observational errors. We interpret them as active
longitudes that appear to migrate in both directions with respect
to the average photometric period but with more-or-less constant
longitudinal separation. There is a short-term rapid migration of
individual spots (appearing as jumps in Fig.~\ref{F7}b, top
panel), e.g. of spot~1 between 1999 and 2000. This migration
during the observing season 2000 was paralleled with a systematic
increase in its area by $\approx$40\% as well as a decay in the
area of spot~2 by even a factor of three from an almost equal area
at the beginning of 2000 to almost disappearance at the end.
During 2000--2004, the separation between the spots increased and
reached a maximum separation in 2004 when indeed a double-humped
light curve was observed. The rapid migration of spot~1 around
2005 caused the two spots to become closer and after that the
separation increased again, reaching an equal maximum in 2008 as
in 2004. From the time-frequency analysis in Fig.~\ref{F3}b, we
find that around 2004, i.e. when the spot longitude separation was
the largest, the photometric period had split into two distinct
values (62\fd0 and 57\fd3). This suggests that in reality the two
spotted regions were at different latitudes on a differentially
rotating surface with $\Delta P/P\approx$0.076 as a lower limit,
albeit a short-term, fast, spot-morphology change could not be
excluded as an alternative explanation. We note that the sign of
the differential rotation cannot be reconstructed because of the
inherent lack of a reliable latitude determination from
one-dimensional photometry. We note again that the longitudinal
separation of the two spots remained nearly constant during the 12
years of observations (150\degr$\pm$17\degr). This suggests that
the two spots, or spotted regions, have a common origin inside the
star and are not just two independent flux tube rings. However, we
see no evidence of a periodic flip flop of the active longitudes,
as found for many other spotted stars with long-term photometric
coverage (e.g. Korhonen \& J\"arvinen \cite{kor:jar}). Only in
1999 and in 2002 was spot~2 larger than spot~1.

\subsection{Modulation of chromospheric tracers}

Our STELLA spectra contain a number of chromospheric activity
tracers, most notably \Halpha \ and Ca\,{\sc ii} H\&K. With a
total time span of four consecutive years, 2006--2010, and a
sampling of the order of one spectrum per (clear) night, the
spectra are suitable for searching a modulation as short as a
fraction of the star's rotational period. Table~\ref{T5}
summarizes the results.

\begin{figure}[tb]
\includegraphics[angle=0,width=8.6cm]{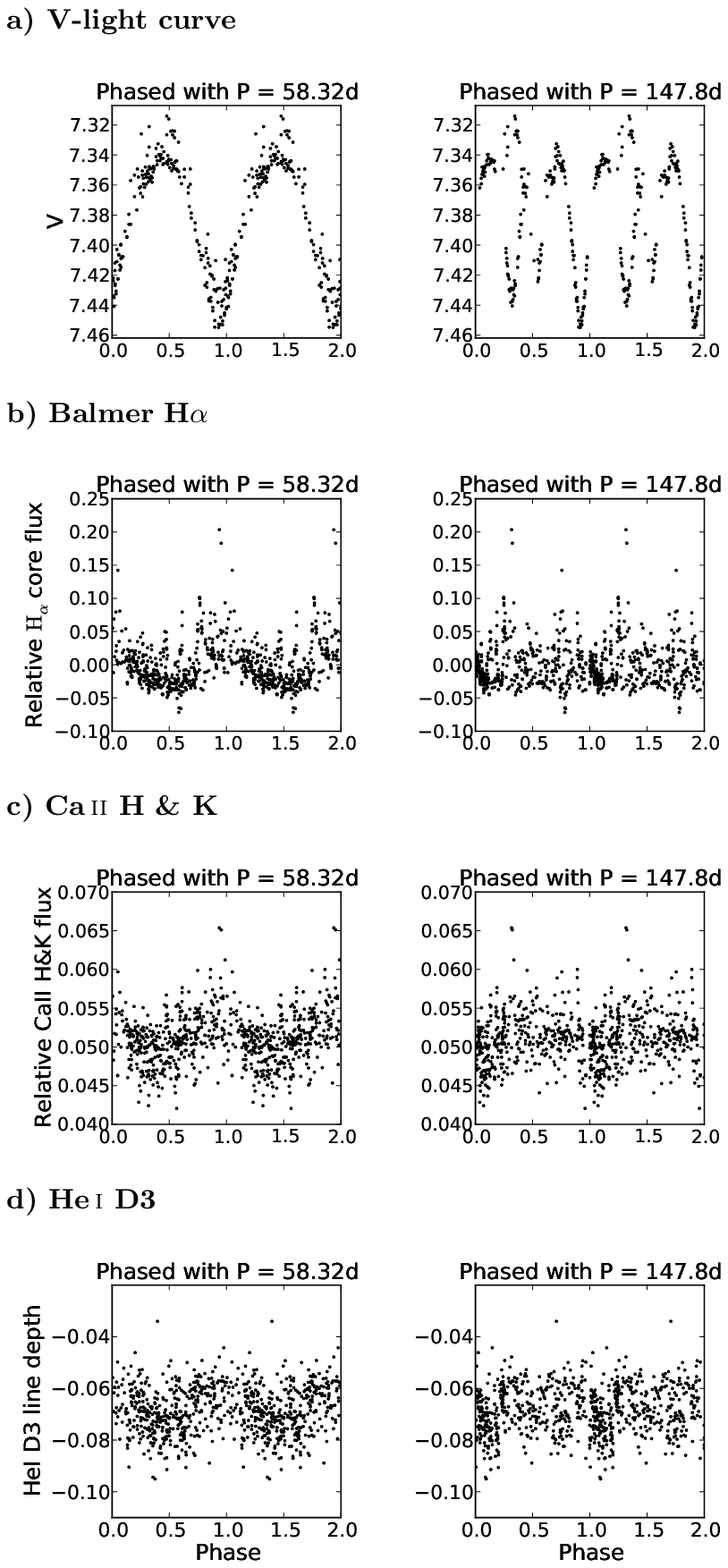}
\caption[ ]{Residual variations for the observing season 2008.
{\bf a} $V$-band light curve, {\bf b} \Halpha -core flux, {\bf c}
relative Ca\,{\sc ii} H\&K emission-line flux, and {\bf d}
He\,{\sc i} D3 absorption line depth. The left column shows the
data phased with the rotational period of 58.32~d, the right
column with the orbital period of 147.89~d, both with respect to a
time of periastron passage. Units are as in Fig.~\ref{F8}. We note
the sharp decline in Ca\,{\sc ii} H\&K flux around periastron in
2008. For better visibility we plot the phases twice.
 \label{F9}}
\end{figure}

\emph{Balmer \Halpha .} The spectra were first re-normalized to
optimize the continuum setting in the echelle order that contains
the \Halpha\ line. We computed a median spectrum from the total of
955 spectra and then iteratively shifted the individual spectra in
intensity and wavelength by minimizing the scatter on both sides
$\pm$0.3nm around the \Halpha\ line. Fig.~\ref{F8}a is a time
series plot of the \Halpha -line profiles for the entire time
series together with a dynamic plot of its residual spectra after
removing the median line profile. We then determined the relative
0.1-nm flux centered on the \Halpha\ laboratory wavelength and
subjected this to a period search with a discrete Fourier
analysis. This search revealed the strongest peak for a period of
58.7$\pm$0.2~d, in agreement with the 58.32-days obtained from our
photometric analysis (note that DFT errors are internal errors).
After prewhitening with this period, two more periods become
significant, one around 90~d and another at the orbital period of
147~d. Because $1/147 + 1/90 \approx 1/58$, we conclude that the
$\approx$90-day peak is an alias and that we see very weak
modulation of \Halpha\ also with the orbital period. The latter is
only marginally significant though with an amplitude of around one
standard deviation. The two panels in Fig.~\ref{F9}b show plots of
the line-core emission phased with the rotational period and the
orbital period, respectively.

\emph{Ca\,\sc{ii} H\&K.} We construct relative Ca\,{\sc ii} H\&K
emission-line fluxes based on the measurement of the flux in a
5.0-nm bandpass centered at 395.0~nm, $F_{50}$, and a 0.1-nm flux
in the two emission lines, $F_{\rm H}$ and $F_{\rm K}$,
respectively. The sum of the emission-line fluxes with respect to
the 5.0-nm flux, $(F_{\rm H}+F_{\rm K})/F_{50}$, as well as the
ratios of each of the two emission lines are then subjected to a
period search similar to \Halpha . Fig.~\ref{F8}b shows dynamic
spectra of both emission-line profiles with a pattern in agreement
with \Halpha , albeit a bit ``cleaner''. The S/N is lower by a
factor of two relative to \Halpha\ because of the lower efficiency
of the SES at blue wavelengths and the star being fainter in the
blue by one magnitude. Nevertheless, the periodogram from the
four-year data set reveals a clear peak at a period of 63$\pm$5
days, which is in moderate but formal agreement with the periods
from \Halpha\ and the optical photometry. Surprisingly, the 90-day
alias and the 147-day peaks are more evident in Ca\,{\sc ii} H\&K
than in \Halpha , although less well-defined for the combined data
in 2007--2010 than for observing season 2008 shown in
Fig.~\ref{F9}c. It indicates variations from year to year. The two
panels in Fig.~\ref{F9}c are phase plots of the relative
emission-line flux (relative to the median profile) with the
rotational and the orbital period, respectively. It shows a
$\approx$1-$\sigma$ drop in flux during periastron passage at
around phase 0 or 1.

\emph{He\,{\sc i} D3.} For this indicator, we chose to simply fit
a Gaussian to its line profile and then use the Gaussian's peak
residual intensity for the period search. Another approach,
similar to measuring the equivalent widths directly between two
defined points in the continuum, did not ensure in the same
homogeneity of the results. This is mostly because the line is
rather weak and narrow with a full-width at the continuum level of
about 0.014~nm, with the red line wing blended by a nearby
Fe\,{\sc i} line (587.630~nm). As we do not resolve the line
width, a Gaussian fit appears to be an acceptable compromise. A
Lomb-Scargle periodogram (e.g. Lomb \cite{lomb}) of the
Gaussian-fit residual intensities from all epochs combined gives a
well-defined peak with a period of 60.68$\pm$0.15~d, i.e. very
close to the expected rotational period of the star. A weighted
wavelet transform analysis (Foster \cite{foster}), i.e. without a
re-sampling of the data along the time axis as for the Lomb
method, shows that this period is present during all seasons.
However, it is always 10-15$\sigma$ longer than the true rotation
period from spot migration. The orbital period does not appear as
clearly as the rotational period. This is partly because of the
shorter exposure time that had been required resolve the rapid
change in the radial velocity during periastron introducing
phase-dependent scatter. Fig.~\ref{F9}d shows the residual
intensity of the line core, again phased with the 58.32-d
rotational period and the orbital period, respectively.

\subsection{Magnetic analysis}

A single STELLA/SES spectrum provides a mean S/N of about 125, too
low for a direct magnetic analysis. However, this can be
compensated for by our having 273 well-sampled spectra in 2008
alone. This time sampling of each individual spectral line profile
can be used in a principal component analysis (PCA;
Bishop~\cite{Bishop95}) approach to boost the S/N of a particular
line by a factor of more than five. It also allows us to search
for phase-dependent variations in the Zeeman-sensitive spectral
line profiles. To differentiate the small excess line-broadening
caused by a surface magnetic field from other line-broadening
mechanisms such as thermal effects, micro- and macroturbulence or
atomic-line parameter variation, e.g. of the oscillator strengths,
we need to use a sophisticated line-profile inversion technique.
Examples of this method that is based on radiative transfer in
magnetic atmospheres has for a long time been applied to various
active cool stars (e.g., Saar et al. \cite{Saar86}, Valenti et al.
\cite{Valenti95}, Ruedi et al. \cite{Ruedi97}, Anderson et al.
\cite{Anderson10}).

\subsubsection{Signal-to-noise ratio enhancement}

A straightforward statistical method that can be readily applied
to individual line profiles is the technique of PCA. The
application of PCA in the present work is twofold. Firstly, it
provides a simple means of detecting the dominant variations in
the data set by building the sample covariance matrix of the
time-sampled measurements for each individual spectral line
profile of interest. This is done in the following separately for
the two spectral lines Fe\,{\sc i}~617.3~nm and FeI\,{\sc
i}~624.0~nm. Calculating its eigenvalues and eigenprofiles allows
us to capture the dominant (energy) variances in the leading
eigenprofiles (principal components). The second motivation for
using PCA is its noise-filtering capability. By retaining only the
leading components in the data, one can obtain a significant
reduction in noise without compromising the basic physical content
in the line profile. Details of the application procedure to
Stokes profiles were described in Carroll et al.
(\cite{Carroll07}) and Mart\'inez et al. (\cite{Martinez08}). We
note that the PCA analysis in this work differs from these studies
in that we build the covariance matrix over the time domain, i.e.
for a particular wavelength (target line) many line profiles
measured over time are used rather than many spectral lines with
different wavelength at a certain time. Applying PCA to our STELLA
spectra and reconstructing the individual line profiles with just
a small number of principal components (three eigenprofiles), we
reached a formal S/N of 650, i.e. an enhancement of approximately
a factor of five with respect to the original data.

\subsubsection{Magnetic flux density inversion}

Our aim is to obtain the magnetic-flux density over the respective
visible surface of the star. The magnetic flux density, or loosely
called magnetic flux, is described by the product $B.f$, the
magnetic field strength $B$ times the surface filling factor $f$.
In the present work no attempt is made to distinguish these two
quantities, our main goal being to provide an estimate of the
magnetic flux and its time variation. All inversions are performed
under the common assumption that the magnetic field is
homogeneously distributed over the surface and radially aligned.

\begin{figure}[tb]
\includegraphics[angle=0,width=6cm]{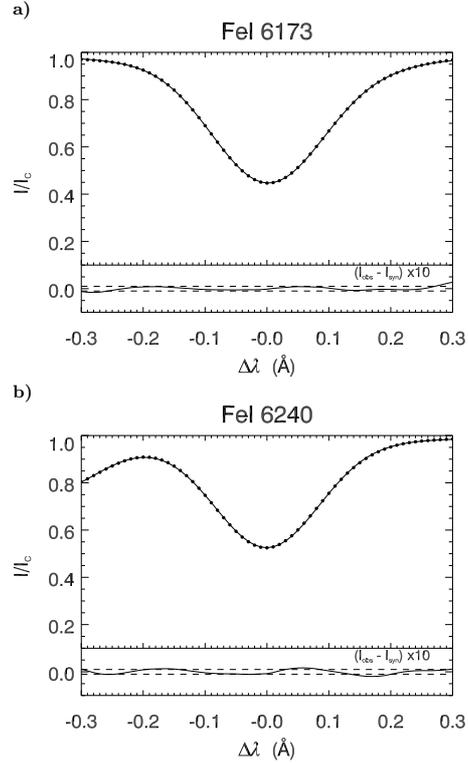}
\caption[ ]{Inversion fits to the averaged line profiles for the
observing season 2008. Panel~{\bf a} shows the Zeeman-sensitive
Fe\,{\sc i}~6173 line (effective Land\'e factor $g_{\rm
eff}$=2.50), panel~{\bf b} shows the Zeeman-insensitive Fe\,{\sc
i}~6240 line ($g_{\rm eff}$=1.00). The dots are the PCA-enhanced
observations, the lines are the inversion fits with a total of six
free parameters. The respective lower panels show the residuals
enhanced by a factor of ten. The average magnetic flux density is
542$\pm$72\,G.
 \label{F10}}
\end{figure}

We follow the analysis of Ruedi et al. (\cite{Ruedi97}) and choose
the Zeeman-sensitive line Fe\,{\sc i}~617.3 (effective Land\'e
factor $g_{\rm eff}$=2.50) and the Zeeman-insensitive line
Fe\,{\sc i}~624.0 ($g_{\rm eff}$=1.00) for a simultaneous analysis
of the two lines. This iron line pair has the advantage of being
of similar strength and having the same excitation potential
($\chi_e$=2.22\,eV) so that their response to temperature
variations can be expected to be similar. Owing to the temperature
inhomogeneities across the stellar surface of HD\,123351, a
simultaneous inversion of this line pair is expected to reduce the
cross-talk between temperature-related broadening and
magnetic-field effects. All spectral line blends found in the VALD
database (Piskunov et al. \cite{Piskunov95}) with a line depth of
more than 0.02 within a 0.03~nm range about the line center of
both iron lines are taken into account.

The flux inversion is done with a variant of our Zeeman-Doppler
imaging code \emph{iMap} (Carroll et al. \cite{Carroll07}). The
forward module of this code consists of an LTE polarized
radiative-transfer computation with the DELO integration method
and is described in more detail in Carroll et al.
(\cite{Carroll08}) to which we refer for more details and
references. In addition to the magnetic flux density $B.f$, we
consider the effective temperature of the atmospheric model as a
free parameter. The atmospheres are Kurucz (\cite{kur}) ATLAS-9
models. Spectral lines are calculated in the L-S coupling regime
and atomic line parameters are taken from VALD, except for the
logarithmic oscillator strength, which is determined by an initial
inversion fit of the average spectral line profile. Assuming the
oscillator strength as a free parameter follows the argument of
Basri \& Marcy (\cite{BM88}) and Ruedi et al. (\cite{Ruedi97})
that solar values are inappropriate for the analysis of different
stellar types. The optimization procedure relies again on the
Levenberg-Marquardt algorithm.

The underlying stellar model used in the inversion consists of a
segmented surface with a $5\degr\times 5\degr$ resolution. Each
segment obtains the magnetic flux density and the effective
temperature of the model atmosphere. During the disk-integration
process, the atmospheric model with its depth stratification of
the temperature and pressure is adjusted according to the
line-of-sight angle to account for limb-darkening in the flux
profiles. In total, the inversion solves for the following six
free parameters for both spectral lines simultaneously; the
magnetic flux density $B.f$, the effective temperature, the
microturbulence, the radial-tangential macroturbulence with equal
radial and tangential components, an instrumental broadening
modeled with a Gaussian velocity distribution, and the oscillator
strength. The rotational velocity and the iron abundance are taken
from the pre-analysis in Sect.~\ref{S3} as they are not expected
to be modulated.

\begin{figure*}
\includegraphics[angle=0,width=\textwidth]{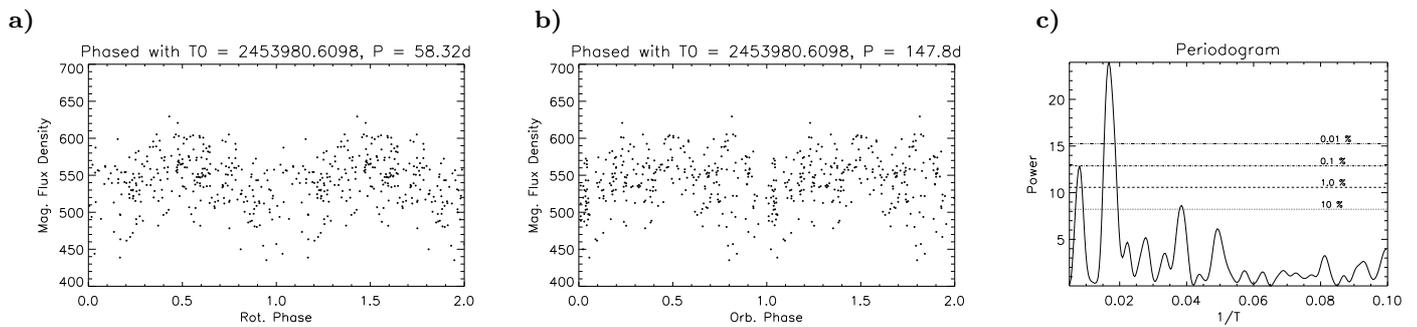}
\caption[ ]{Magnetic flux variation for the observing season 2008.
{\bf a} Phase plot with the stellar rotation period of 58.32~d,
{\bf b} phase plot with the orbital period of 147.89~d, and {\bf
c} Lomb-Scargle periodogram from the 273 individual spectral line
inversions. The two strongest peaks in the periodogram agree with
the rotational and the orbital periods, respectively. The
strongest peak refers to a period of 58.7~d, the second strongest
peak to 144.3~d. Note again that we plot the phases twice for
clearer visibility.
 \label{F11}}
\end{figure*}

\subsubsection{Results from the inversion}

We concentrate on the data set obtained in the year 2008, as was
the most finely sampled. Selecting only the highest S/N spectra
still gives a database of 273 spectral line profiles for each of
the above iron lines. In the first step, we only invert the
average line profiles (zeroth eigenprofile) that determine the
global surface-averaged values of $B.f$ and the other free
parameters. This gives a first hint on the average strength of
this field. In addition, we can determine the other free
parameters and then keep them fixed for the following analysis of
the individual line profiles.

From Fig.~\ref{F10}, we see that the fits to the averaged line
profiles are extremely precise and consistent with the low noise
level and a reduced $\chi^2$ error of 1.12 for the Fe\,{\sc
i}~617.3 line, and 1.58 for the Fe\,{\sc i}~624.0 line. The
obtained (logarithmic) oscillator strengths are --2.78 for the
Fe\,{\sc i}-617.3 line, and --3.18 for the Fe\,{\sc i}-624.0 line.
Micro- and macroturbulence were found to be 1.15~\kms\ and
3.9~\kms , respectively, and the average effective temperature to
4830~K (with a peak-to-valley variation of 60~K). The inferred
magnetic flux density was 542\,G with a formal error of 72\,G
calculated from the covariance matrix (see Carroll et al.
\cite{Carroll07}). We emphasize that the continuum depression
caused by a slightly reduced effective temperature related in turn
to a non-zero spot-filling factor, is implicitly solved for in our
inversion. If we plot the effective temperature of each line pair
versus rotational (or orbital) phase, it shows plots similar to
those in Figs.~\ref{F11}a and b. This is expected if the magnetic
flux density stems from localized regions rather than from a
global network. An indication of the reality of the magnetic field
broadening is obtained from a number of inversions without a
magnetic field. In all of these zero-field cases, the quality of
the fits ($\chi^2$ of 1.74 and 1.92 for the two lines,
respectively) were unable to reach that of the magnetic case. This
is also obvious in the projected $\chi^2$ landscape, which shows
that the zero-field case is far beyond the 5-$\sigma$ region of
the $\chi^2$ minimum with a magnetic field. Therefore, we consider
our magnetic-flux result as significant.

The next step in our analysis is to invert the 273 individual line
profiles of our iron-line pair to search for a systematic periodic
behavior. Fig.~\ref{F11} shows the results and indicates a clear
rotational dependence on the magnetic flux, as well as a (weaker)
dependence on the orbital phase. The individual formal errors in
the retrieved magnetic flux values are between 55 and 80~G,
depending on the initial S/N. We note again that by performing a
simultaneous inversion of the two iron lines with both having a
similar temperature response but a different magnetic response, we
reduce the crosstalk of temperature effects and its possible
misinterpretation in the inversion process. A Lomb-Scargle
periodogram (Fig.~\ref{F11}c) indeed indicates a pronounced peak
above a 0.01\% false-alarm probability at 58.7~d with an
uncertainty of 4.3~d determined from a Monte Carlo resampling. A
second significant peak, but not quite with as high a probability
as the first peak, shows a periodicity of 144.3~d with an
uncertainty of 8.9~d. The distribution of the summed squared
errors for each line-profile fit versus phase does not indicate
any systematic distribution, suggesting that the inversion was not
fooled by a systematic, residual surface-temperature variation. As
in Fig.~\ref{F9} for the chromospheric indicators, we plot the
magnetic flux densities over the rotational and orbital phase in
Fig.~\ref{F11}ab.

\section{Discussion}\label{S5}

\subsection{Circularization and synchronization}

Tidal interaction between stars can couple their spin to their
orbital angular momentum such that dissipation of tidal energy
causes a binary to circularize its orbit (e.g. Zahn \cite{z89} and
earlier papers; Tassoul \& Tassoul \cite{tt96}). Observational
results indicate that most active binaries have circular orbits
with synchronized rotation and a spin axis perpendicular to the
orbital plane (e.g. Fekel \& Eitter \cite{fe89}). Other systems
with weaker tidal interactions may not be currently old enough to
have reached complete synchronization or circularization. However,
these conditions may occur once tidal interactions have operated
long enough or after a system evolved to a state in which tidal
interactions are enhanced. For late-type stars with convective
envelopes, Zahn (\cite{z89}) investigated the effects of the
equilibrium tide on synchronization and circularization, while
Tassoul (\cite{t88}) explored the theory of geometrical
distortions that cause large-scale hydrodynamic currents. Although
these two theories disagree significantly on absolute timescales,
both predict that synchronization should occur before
circularization.

HD~123351 is neither synchronized nor circularized. Its large
orbital eccentricity suggests a pseudo-synchronous rotation period
of $\approx$12~d according to the tidal-friction theory of Hut
(\cite{hut}), which is just 8\% of the orbital period. With a
58.3-d rotation period, the star appears to be a strongly
asynchronous and slow rotator. Since all convective stars likely
undergo some sort of differential surface rotation, one would
expect a certain range of rotation periods if derived from
photometric spot activity because spots could appear at various
stellar latitudes. For comparison, the (strong) solar differential
rotation allows a spread of rotation periods of $\approx$10\%.
Comparable results are found for other active stars based on
Doppler imaging (e.g. Barnes et al. \cite{barnes}, Strassmeier
\cite{aarev}). However, the difference between the observed
photometric period and the pseudo-synchronous rotation period for
HD~123351 is more than 500\% and could not be explained by
differential surface rotation.

Therefore, we ask what caused a cool star to fail to reach
equilibrium in terms of its spin and orbital angular momentum
after 6--7~Gyr? Although not in disagreement with the prediction
that synchronization shall occur before circularization, the
Tassoul and Zahn mechanisms predict vastly different timescales
for HD~123351. While the original Zahn (\cite{z89}) equations
predict synchronization and circularization well beyond the age of
the universe, the Tassoul (\cite{t88}) equations predict
synchronization after $\approx$200,000 years and circularization
after $\approx$3~Gyr. Both can obviously be excluded because of
the measurements derived from the observations. Additional
mechanisms were proposed to account for the strong tidal coupling
required by the observations of circularization of binary systems
in open clusters of various age (Ogilvie \& Lin \cite{ogi:lin}),
but we refer to the refutation of Tassoul's hydrodynamical
mechanism by the proper accounting of an Ekman layer by Rieutord
\& Zahn (\cite{rie:zah}). In the case of HD~123351, the orbital
and rotational frequencies today imply that inertial oscillations
(waves) might have neen excited in the star leading to an average
tidal dissipation approximately 10--100 times that predicted by
Zahn's theory. Such inertial waves are not described in the
traditional approximations and correspondingly reduce the
estimated synchronization and circularization times. Because
non-linearities are expected for eccentric binaries and for
components older than the Sun, no simple prediction of Ogilvie \&
Lin (\cite{ogi:lin}) is possible for HD~123351, but in general
brings the theory closer to the observation.

\subsection{Inter-binary activity?}

The assumption of an inter-binary magnetic field that could
transfer orbital to spin momentum and vice versa may indeed be
inaccurate for intervals longer than purely gravitationally and
hydrodynamically predicted timescales. It is well known that
inter-binary magnetic fields focus the accretion stream in
cataclysmic variables (CV) containing a magnetic white dwarf (e.g.
Warner \& Woudt \cite{war:wou}), as well as some Algol systems
(e.g. Richards \cite{rich}). Reconnection processes within these
fields may also act as the source of inter-binary flares,
suggested e.g. for the RS~CVn binary system HR~5110 (Graffagnino
et al.~\cite{hr5110}). The field lines in general provide a
channel for the exchange of charged particles between the two
stellar components. Its principal existence on a geometrically
shorter scale is nicely demonstrated by the Jupiter-Io system
(e.g. Su \cite{su}). However, the closest distance between the two
components in the HD~123351 binary system ($\approx$5~R$_\star$)
is still larger than for the aforementioned CV and Algol binaries,
thus the expected field density may be insufficient for
inter-binary flares. No such superflare was detected in our data.
Coronal mass ejections may however still exist, which would have
had an effect on the evolution of the spin and orbital angular
momentum of the star but remains speculative.

\subsection{Orbital-induced magnetic activity?}

In a very eccentric close binary similar to HD\,123351, we might
have the rare possibility of exploring a hypothetical efficiency
increase and decrease of a stellar dynamo when the two components
are closest and farthest, respectively, assuming e.g. that some
sort of extra convective mixing couples the dynamo action to tidal
forces. Glebocki \& Stawikowski (\cite{gle:sta1}, \cite{gle:sta2})
found some relation between chromospheric Ca\,{\sc ii} emissions
and the orbital parameters of late-type giants and suggested that
tidal forces are somehow responsible for the enhanced
chromospheric activity. In contrast, Basri (\cite{basri87}) ruled
out tidal coupling as an explanation of the increased activity in
binaries. Schrijver \& Zwaan (\cite{sch:zwa}) argued that in a
synchronized close binary the relevant axis for rotational effects
is through the center of gravity of the binary system and thereby
could affect the stellar dynamo in such a way that it produces
enhanced magnetic activity relative to single stars. Zaqarashvili
et al. (\cite{zaq}) assumed that the deviation from a spherical
shape of a binary component due to tidal interaction excites a
fundamental pulsation mode that then amplifies the torsional waves
in the interior magnetic field and eventually causes greater
surface activity. It has been found that (extra-solar) planets can
induce chromospheric and coronal activity in the form of a hot
spot (Shkolnik et al. \cite{shk}), although not all stars with
close-in planets are prone to these star-planet interactions (e.g.
Fares et al. \cite{fares}). Lanza (\cite{lanza08}, \cite{lanza09})
argued that a flux rope between the host star's coronal magnetic
field and the planet's magnetosphere experiences a decrease in
relative helicity and reacts with an energy release from the flux
rope's footpoint(s). His model predicts a greater prominence
activity for stars with planets than without, caused by the
accumulation of matter evaporated from the planet. A very similar
picture may be adopted for binaries such as HD~123351 that have a
low-mass dwarf star as a companion instead of a close-in hot
Jupiter.

Our Ca\,{\sc ii} H\&K and He\,{\sc i} data indeed indicate a
correspondence of magnetic activity with orbital periastron, at
least during observing season 2008. No clear relation was seen for
the observing seasons 2007 and 2009. However, this may be caused
by some bias because the season 2008 was the one with the highest
data quality and finest data sampling, both of which are needed to
detect such a correspondence. Furthermore, the modulation
amplitudes were always small and on the order of the annual rms.
But the detection is quite surprising because \Halpha\ does not
show a similar correspondence at all, which is the opposite of
what we would expect if there is prominence activity (e.g. Mackay
et al. \cite{mack}, Frasca et al. \cite{frasca}). Our
magnetic-flux analysis also revealed evidence of a weak
orbital-phase coherence in 2008 and thus strengthens the Ca\,{\sc
ii} and He\,{\sc i} detections. The $B.f$ orbital variation is in
phase with the variation in both Ca\,{\sc ii} and He\,{\sc i} in
the sense that lower chromospheric emission is seen when the
(photospheric) magnetic flux density is lower, with their
respective minima coinciding with a time when the two binary
components are closest in distance, i.e. at periastron. The
orbitally phased $V$-band light curve in Fig.~\ref{F9}a is
completely out of phase, as expected because it is purely
photospheric in origin. The comparison of the chromospheric
emissions with the magnetic flux density is more puzzling when
plotted with the rotational period. Then, $B.f$ varies exactly in
anti-phase with all three chromospheric lines, in particular also
with \Halpha\ (compare the left columns in Fig.~\ref{F9}bcd with
Fig.~\ref{F11}a), which is unexpected. The rotationally phased
$V$-light curve appears nicely in phase with $B.f$ and in anti
phase with the chromospheric lines, as expected if chromospheric
plages are superposed on photospheric spots as in the Sun. In
summary, we take the anti-phase relation of the magnetic flux with
the chromospheric emissions as evidence that there are two
magnetic fields present at the same time, a localized surface
magnetic field associated with spots and a global field that is
oriented towards the secondary component such as an inter-binary
field. We have no already developed model to combine all of this
but conclude that the $B$-component that we measure in $B.f$
possibly stems from a mixture of bright plage-like regions and
dark spots, both with different respective filling factors ($f$),
but the former also caused by the interaction with the secondary
star.

The overall amplitude of the radial-velocity residuals shown in
Fig.~\ref{F2}b suggests that the star was more asymmetrically
spotted in 2007 than in 2008, albeit the $V$-light curves showed
approximately the same amplitudes and shapes. In 2007, the \Halpha
-core flux indeed showed the most pronounced variations
($\approx$4$\sigma$) of all seasons when phased with the rotation
period but again basically just a scatter plot when phased with
the orbital period (not shown). This indicates that \Halpha\
responds to the same surface inhomogeneities as seen in the other
lines or the photometry. It encourages us to believe that the
\Halpha -profile variations of HD\,123351 originate predominantly
from near the stellar surface and have no detectable circumstellar
(absorption) component, as would be expected for coronal
prominence activity. We note that we did not apply a correction
for the photospheric contribution to \Halpha\ but did all
time-series analysis from residual line profiles after subtraction
of a median profile.

\section{Summary and conclusions}\label{S6}

We have obtained a four-year-long, phase-resolved, spectral
database of the active star HD\,123351 using our new robotic
telescope facility STELLA in Tenerife. These data are complemented
with 12 years of photometric monitoring with our automatic
photometric telescope {\sl Amadeus} in Arizona. Our summary and
conclusions are as follows:

\begin{itemize}
\item We have discovered HD\,123351 to be a single-lined spectroscopic binary
with a period of 147.8919~d and an orbital eccentricity of 0.809.
\item The STELLA Echelle Spectrograph enabled the most precise (and
presumably also most accurate) orbit determination of any active
binary so far in the literature with an rms of the residuals of just
47~\ms .
\item A period of $\approx$60~d was obtained from 12 years of {\sl Amadeus}
$VI$-band photometry and has been converted into a true rotation
period of 58.32~d from a linear fit to the two dominant
spot-migration curves. The period difference from the two spots in
2004 suggests a lower limit to the surface differential rotation
of $\Delta P/P$ of 0.076, which is roughly half the solar value.
\item The global rotation period of HD\,123351 was also recovered from the
orbital radial-velocity residuals.
\item A spectrum synthesis yielded average stellar atmospheric parameters
of $T_{\rm eff}$=4780$\pm$70~K, $\log
g$=3.25$\pm$0.30, and [Fe/H]=$0.00\pm0.08$. The Hipparcos parallax
in combination with evolutionary tracks suggests a mass of
1.2$\pm$0.1~M$_\odot$ and an age of 6-7~Gyr, consistent with the
measured $^{12}$C/$^{13}$C ratio of $\approx$40.
\item The lithium abundance of $\log n$=1.70 appears higher than
expected for that age but the line also appears broader than
expected for the $v\sin i$ of 1.8$\pm$0.7~\kms . We suspect that
its asymmetry is partly due to a significant contribution from the
$^6$Li isotope.
\item From an application of the K\"uker-Stix differential-rotation
code to a MESA stellar model with matching parameters of
HD\,123351, we obtain a convective turn-over time that spans
between a maximum of 80~d near the bottom of the convection zone
to 10~d near the surface. This implies an average Rossby number
greater than unity.
\item Our light-curve analysis has reconstructed two persistent spotted
regions with a nearly constant longitudinal separation of
150\degr$\pm$17\degr\ over the course of 12 years. The spot
temperatures were on average 1160~K lower than the effective
temperature covering between 5--10\% of the entire surface. The
photometry showed no signs of a flip flop nor of flares.
\item Chromospheric emission-line fluxes of $\approx$10$^6$
erg\,cm$^{-2}$s$^{-1}$ were measured for Ca\,{\sc ii} H\&K, \Halpha
, and the Ca\,{\sc ii} infrared triplet. The detection of the
He\,{\sc i} D3-line at 587.56~nm also indicated the presence of
non-thermal heating processes.
\item Time-series analysis of above chromospheric activity indicators
yielded a clear modulation with the stellar rotation period but
also provided evidence of a modulation with the orbital period,
albeit less well-defined.
\item We measured the Zeeman broadening from a pair of
iron lines and obtained an average surface magnetic flux density
of 542$\pm$72~G. A time-series analysis of the data from 2008
yielded again a clear modulation with the rotation period and a
weak modulation with the orbital period in the sense that a lower
magnetic flux was seen at times near periastron.
\item It remains unclear why the magnetic
flux varies exactly in anti-phase with the chromospheric emissions
if plotted versus the rotational period, and is lowest near
periastron when plotted versus the orbital period. The measured
magnetic flux possibly stems not only from cool photospheric spots
but also from bright plage-like regions.
\item We postulate an inter-binary magnetic field and tentatively
conclude that it is responsible for the flux dilution at
periastron. It is also possibly responsible for the unexpected
slow and asynchronous rotation of the primary star.
\end{itemize}

\acknowledgements{STELLA was made possible by funding through the
State of Brandenburg (MWFK) and the German Federal Ministry of
Education and Research (BMBF). The facility is a collaboration of
the AIP in Brandenburg with the IAC in Tenerife. We thank all
engineers and technicians involved, in particular Manfred Woche
and Emil Popow and his team, and Ignacio del Rosario and Miguel
Serra from the IAC Tenerife day-time crew. We also thank M.
K\"uker for computing the convective turn-over times for us. KO
thanks Z. Kollath for useful discussions and acknowledges support
from the Hungarian research grant OTKA K-081421. Finally, we thank
the referee, Dr. Antonio Lanza, for a very constructive report
that helped to improve the paper considerably. }

\end{document}